\documentclass[10pt, letter, onecolumn]{arxiv}
\usepackage[utf8]{inputenc}
\usepackage{textgreek}
\usepackage{kantlipsum, lipsum}
\usepackage{dm-colors}
\usepackage{xcolor}
\usepackage{makecell}
\usepackage{amsmath}
\usepackage[switch]{lineno}
\usepackage{pstricks, pst-node}
\usepackage{verbatim}
\usepackage{multirow}
\usepackage{rotating}
\usepackage{scalerel}
\usepackage{booktabs}
\usepackage{enumitem}
\usepackage{xspace}
\usepackage{comment}
\usepackage[breakable]{tcolorbox}
\usepackage{bm}
\usepackage{bbm} 
\usepackage{mathtools}
\usepackage{soul}
\usepackage{epsfig}
\usepackage{graphicx}
\usepackage{amssymb}
\usepackage{colortbl}
\usepackage{csquotes}
\usepackage{setspace}
\usepackage{colortbl}
\usepackage{bbding}
\usepackage{threeparttable}
\usepackage{tabularx,ragged2e}
\usepackage{placeins}
\usepackage{bbding}
\usepackage{algorithm}
\usepackage{algorithmic}
\usepackage{ulem}

\definecolor{darkpastelgreen}{rgb}{0.13, 0.55, 0.13}
\definecolor{darkpastelred}{rgb}{0.55, 0.13, 0.13}
\usepackage[hang,flushmargin]{footmisc}

\usepackage{nameref}
\usepackage{varioref}
\usepackage{amssymb}%
\usepackage{lineno}
\usepackage{pifont}
\usepackage{xcolor}

\usepackage[pagebackref,breaklinks,colorlinks,
    citecolor=blue,
    filecolor=magenta,
    linkcolor=red,
    urlcolor=cyan]{hyperref}
    
\usepackage[noabbrev,capitalize]
{cleveref}
\usepackage{etoc}

\definecolor{reasoncolor}{HTML}{EE822F}
\definecolor{diagnosecolor}{HTML}{C81D31}
\definecolor{lookupcolor}{HTML}{1E5799}
\definecolor{matchcolor}{HTML}{3B82C4}
\definecolor{searchcolor}{HTML}{5BA3E0}
\graphicspath{{figures/}}
\definecolor{midgreen}{rgb}{0.0, 0.75, 0.0}
\definecolor{softblue}{HTML}{136783}

\newcommand{\jc}[1]{\textcolor{black}{#1}}

\newcommand{\figsubref}[2]{Fig.~\hyperref[#1]{\ref*{#1}#2}}

\definecolor{goodgreen}{HTML}{00B050}
\definecolor{badred}{HTML}{FF0000}

\title{\Large{Predicting Neuromodulation Outcome for Parkinson's Disease with Generative Virtual Brain Model}}

\author[1,4]{Siyuan Du$^\dagger$} 
\author[4,6]{Siyi Li$^\dagger$} 
\author[5]{Shuwei Bai}
\author[5]{Ang Li}
\author[1,4]{Haolin Li}
\author[8]{Mingqing Xiao}
\author[9]{Yang Pan}
\author[8]{Dongsheng Li}
\author[3,4]{Weidi Xie}
\author[3]{Yanfeng Wang}
\author[3,4,7]{Ya Zhang$^\ast$}
\author[5]{Chencheng Zhang$^\ast$}
\author[2]{Jiangchao Yao$^\ast$}

\affil[1]{\normalsize College of Computer Science and Artificial Intelligence, Fudan University, Shanghai, China \authorcr  \vspace{0.1cm}}
\affil[2]{\normalsize Cooperative Medianet Innovation Center, Shanghai Jiao Tong University, Shanghai, China \authorcr \vspace{0.1cm}}
\affil[3]{\normalsize School of Artificial Intelligence, Shanghai Jiao Tong University, Shanghai, China \authorcr \vspace{0.1cm}}
\affil[4]{\normalsize Shanghai AI Laboratory, Shanghai, China \authorcr \vspace{0.1cm}}
\affil[5]{\normalsize Ruijin Hospital, Shanghai Jiao Tong University School of Medicine, Shanghai, China\authorcr \vspace{0.1cm}}
\affil[6]{\normalsize University of Science and Technology of China, Anhui, China \authorcr \vspace{0.1cm}}
\affil[7]{\normalsize Institute of Artificial Intelligence for Medicine, Shanghai Jiao Tong University School of Medicine, Shanghai, China\authorcr \vspace{0.1cm}}
\affil[8]{\normalsize Microsoft Research Asia, Shanghai, China \authorcr \vspace{0.1cm}}
\affil[9]{\normalsize Zhongnan Hospital of Wuhan University, Hubei, China \authorcr \vspace{-0.1cm}}

\renewcommand{\correspondingauthor}[1]{
$^\dagger$~These authors contributed equally. 
$^\ast$~Corresponding author. Email addresses: sunarker@sjtu.edu.cn (J. Yao)}

\begin{document}

\begin{abstract}
Parkinson's disease~(PD) affects over ten million people worldwide.
%
%
Although temporal interference (TI) and deep brain stimulation (DBS) are promising therapies, inter-individual variability limits empirical treatment selection, increasing non-negligible surgical risk and cost.
%
%
Previous explorations either resort to limited statistical biomarkers that are insufficient to characterize variability, or employ AI-driven methods which is prone to overfitting and opacity.
%
%
We bridge this gap with a pretraining-finetuning framework to predict outcomes directly from resting-state fMRI.
%
%
Critically, a generative virtual brain foundation model, pretrained on a collective dataset (2707 subjects, 5621 sessions) to capture universal disorder patterns,
%
%
was finetuned on PD cohorts receiving TI (n=51) or DBS (n=55) to yield individualized virtual brains with high fidelity to empirical functional connectivity (r=0.935). 
By constructing counterfactual estimations between pathological and healthy neural states within these personalized models, we predicted clinical responses (TI: AUPR=0.853; DBS: AUPR=0.915), substantially outperforming baselines.
%
%
External and prospective validations (n=14, n=11) highlight the feasibility of clinical translation.
Moreover, our framework provides state-dependent regional patterns linked to response, offering hypothesis-generating mechanistic insights. 
%

\end{abstract}

\maketitle
%

\section{INTRODUCTION}
\label{sec:1-intro}

\jc{Parkinson's disease (PD) is a multifactorial worldwide neurodegenerative disorder, which affects over 10 million individuals and induces a substantial healthcare burden~\cite{luo2025global}. For its treatment, pharmacotherapy is generally effective in the early management of progressive motor symptoms, but treatment-limiting complications often develop with disease progression~\cite{freitas2017motor}.} 
\jc{Neuromodulation therapies have therefore been considered as promising alternatives during the moderate to advanced stages of PD.}
\jc{Specifically, two typical neuromodulation modalities include deep brain stimulation (DBS) and transcranial temporal interference stimulation (TI), where the former is an established invasive therapy that delivers chronic electrical stimulation via implanted electrodes to improve motor function~\cite{deep2001deep}, whereas the latter is a non-invasive approach that seeks to modulate deep brain circuits by exploiting interference envelopes generated by multiple high-frequency currents~\cite{demchenko2025human}.}
As patients move toward neuromodulation, treatment planning becomes consequential and clinically sensitive.

\begin{figure}[t!]
\vspace{-20pt}
  \centering
  \includegraphics[width=0.95\linewidth]{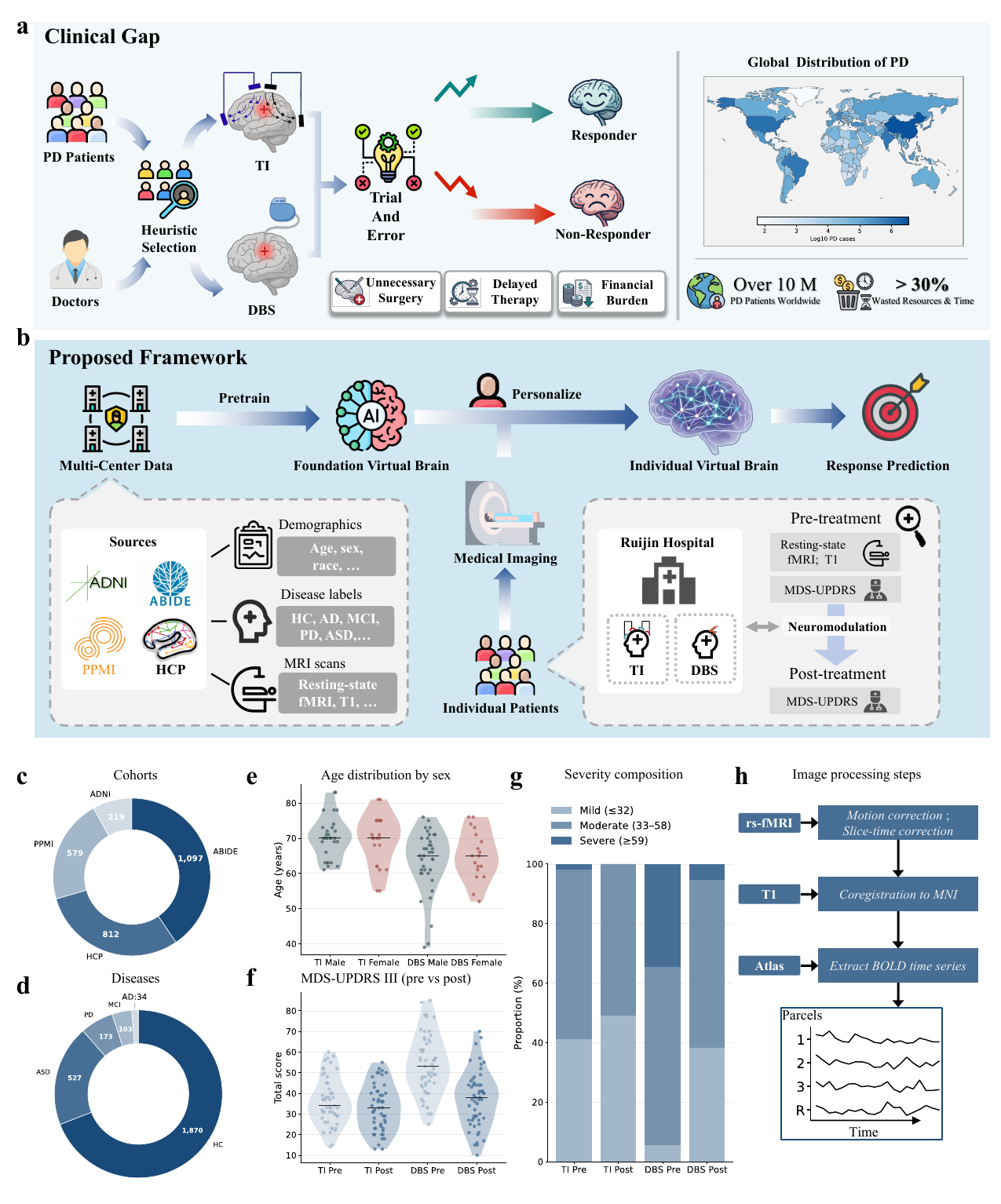} 
  \caption{\textbf{Clinical gap, proposed framework, and cohort statistics.} 
  \textbf{a,} Overview of trial-and-error neuromodulation \jc{process for PD and the clinical situation}.
\textbf{b,} Overview of the proposed framework: a generative \jc{virtual brain paradigm} transfers dynamical priors from large-scale data to small clinical cohorts for individualized neuromodulation response prediction. 
\textbf{c--d,} \jc{Statistics} of the pretraining cohort: sample distribution across four centers (\textbf{c}) and disease spectrum composition (\textbf{d}).
\textbf{e--g,} \jc{Statistics} of the Ruijin PD cohorts: age distribution stratified by sex (\textbf{e}), MDS-UPDRS III motor scores distribution (\textbf{f}) and PD severity composition (\textbf{g}).
\textbf{h,} Image processing pipeline: from raw BOLD acquisition to parcellated regional time series.
}
\vspace{-20pt}
\label{fig:intro}
\end{figure}

\jc{However, the existing selection criteria to adopt neuromodulation heavily rely on the neuroscientific experiences regarding demographic factors, disease duration, and dopaminergic responsiveness~\cite{dallapiazza2018considerations},}
while the treatment response to both DBS and TI exhibits substantial inter-individual variability~\cite{hariz2022deep}.
For DBS, even among rigorously screened candidates, approximately 30--40\% of patients fail to achieve clinically meaningful motor improvement~\cite{somaa2021deep}.
For TI, \jc{although early studies likewise report heterogeneous outcomes, there is a lack of generally effective criteria to support patient stratification~\cite{demchenko2025human}.} 
%
\jc{This absence of reliable pre-treatment estimates often leaves patients facing prolonged trial-and-error before an effective therapy is identified.}
\jc{In turn, overestimation may prompt unnecessary invasive surgery with added costs and hardware-related complications (e.g., intracranial hemorrhage, infection, and lead failure)~\cite{rasiah2023complications}, whereas underestimation can delay effective treatment and potentially forfeit an optimal window for symptom control~\cite{desouza2013timing} as shown in \figsubref{fig:intro}{a}.}

%

To advance the estimation, previous clinical practice has derived digital biomarkers for PD from neuroimaging 
at the group level~\cite{younce2021resting,horn2017connectivity}.
\jc{For example,} structural markers (e.g., cortical thickness, subcortical volumes) and tract-level features have been investigated for DBS planning and outcome association~\cite{horn2017connectivity,muthuraman2017effects}, but findings are often heterogeneous across cohorts and pipelines~\cite{andrews2023structural}. 
In parallel, resting-state fMRI has enabled connectivity-based biomarkers, where network-level markers are extracted to characterize PD-related circuit dysfunction and to correlate with neuromodulation benefit~\cite{younce2021resting, wang2021normative}. 
Nevertheless, these group-level biomarkers often fail to generalize well to individual patients due to inter-individual variability~\cite{tahmasian2017resting}.

\jc{For AI-driven methods, they seek to decode individual network states directly from resting-state fMRI, including end-to-end learning of functional connectivity~\cite{riaz2017fcnet,li2021braingnn} and transformer-based representation learning for brain dynamics~\cite{caro2023brainlm}.}
However, there are two-fold challenges remaining: 1) fMRI signals are high-dimensional, sparse, and confounded by site/subject-specific noise~\cite{murphy2013resting,noble2017multisite};  \jc{2) labeled PD neuromodulation cohorts (especially for TI) are severely limited, as the collection and annotation require a well-characterized setting.}
Consequently, naive AI modeling on small datasets is prone to overfitting and lacks mechanistic interpretability~\cite{arbabshirani2017single,avbervsek2022deep}. 
\jc{The difficulty presented in this area limits the advances of AI-aided PD therapy.}
%
%

\jc{These gaps motivate us to explore a more generalizable framework for precision neuromodulation. Specifically, we are inspired by the remarkable success of foundation models for medicine, which are first pretrained on large-scale data to learn broad knowledge, and then achieve impressive adaptation performance with only a few samples in the downstream scenarios~\cite{bommasani2021opportunities,moor2023foundation,singhal2023large}.
In this study, we first collect a large-scale open-source rs-fMRI dataset encompassing a range of neurodegenerative disorders. Then, a generative foundation virtual brain model (FVB) that fits the BOLD signal translated from rs-fMRI is designed to learn the whole-brain hemodynamics. By personalizing FVB with the data of each patient, we instantiate an individualized virtual brain model (iVB) that breaks the inter-individual variability dilemma and accurately characterizes the brain dynamics for PD, which is illustrated in \figsubref{fig:intro}{b}.}
\jc{Furthermore, we introduce novel Counterfactual Brain Mismatch (CBM) features as mechanistically interpretable digital biomarkers for the neuromodulation outcome prediction with the appealing merit of iVBs: iVBs enable counterfactual simulations that quantify patient-specific deviations from normative dynamics (e.g., ``what-if-healthy'' and ``what-if-distorted'' contrasts). Notably,} these deviations are intrinsically state-dependent and circuit-resolved, supporting not only prediction but also identification of state-dependent circuit signatures that explain why stimulation succeeds or fails.

%
\jc{We perform extensive experiments to demonstrate the effectiveness of our method.}
Across both non-invasive and invasive stimulation (TI, n = 51; DBS, n = 55; cohort statistics are provided in \figsubref{fig:intro}{e--g}), the framework accurately discriminates responders from non-responders, significantly outperforming both conventional neuroimaging biomarkers and AI-driven methods.
Crucially, the framework shows encouraging performance at an external center without site‑specific retraining and remains feasible in a small prospective setting.
Meanwhile, predictions arise from explicit, region-resolved deviations from normative dynamics, allowing clinicians to trace outcomes back to circuit-level states rather than opaque algorithmic scores.
Beyond forecasting, this capacity enables hypothesis-generating, mechanism-oriented analyses by nominating state-dependent circuit signatures of benefit, yielding interpretable insights for both mechanistically opaque TI and established DBS.
Together, our results provide evidence that generative virtual-brain modeling is a practical route toward data-efficient, explainable, and mechanism-informed treatment selection, advancing neuromodulation beyond heuristic trial-and-error toward true precision care.

\section{RESULTS}
\jc{The resting-state fMRI dataset to pretrain the generative foundation virtual brain model contains} 5,621 sessions from 2,707 unique participants across four public cohorts: Alzheimer’s Disease Neuroimaging Initiative (ADNI)~\cite{jack2008alzheimer},  Parkinson's Progression Markers Initiative (PPMI)~\cite{marek2011parkinson},  Autism Brain Imaging Data Exchange (ABIDE)~\cite{di2014autism}, and Human Connectome Project (HCP)~\cite{van2013wu}. \jc{The statistics have been illustrated in \figsubref{fig:intro}{c--d}.}
\jc{The pretrained FVB was then finetuned on in-hospital} pre-treatment fMRI data from 106 Parkinson’s disease patients (51 transcranial interference [TI] recipients and 55 deep brain stimulation [DBS] recipients) 
to \jc{acquire individualized virtual brain models (iVBs) respectively. Using counterfactual simulation features computed from iVBs, we performed treatment outcome predictions} and extracted state-dependent biomarkers. 



\begin{figure}[htbp]
\vspace{-20pt}
  \centering
\includegraphics[width=0.95\linewidth]{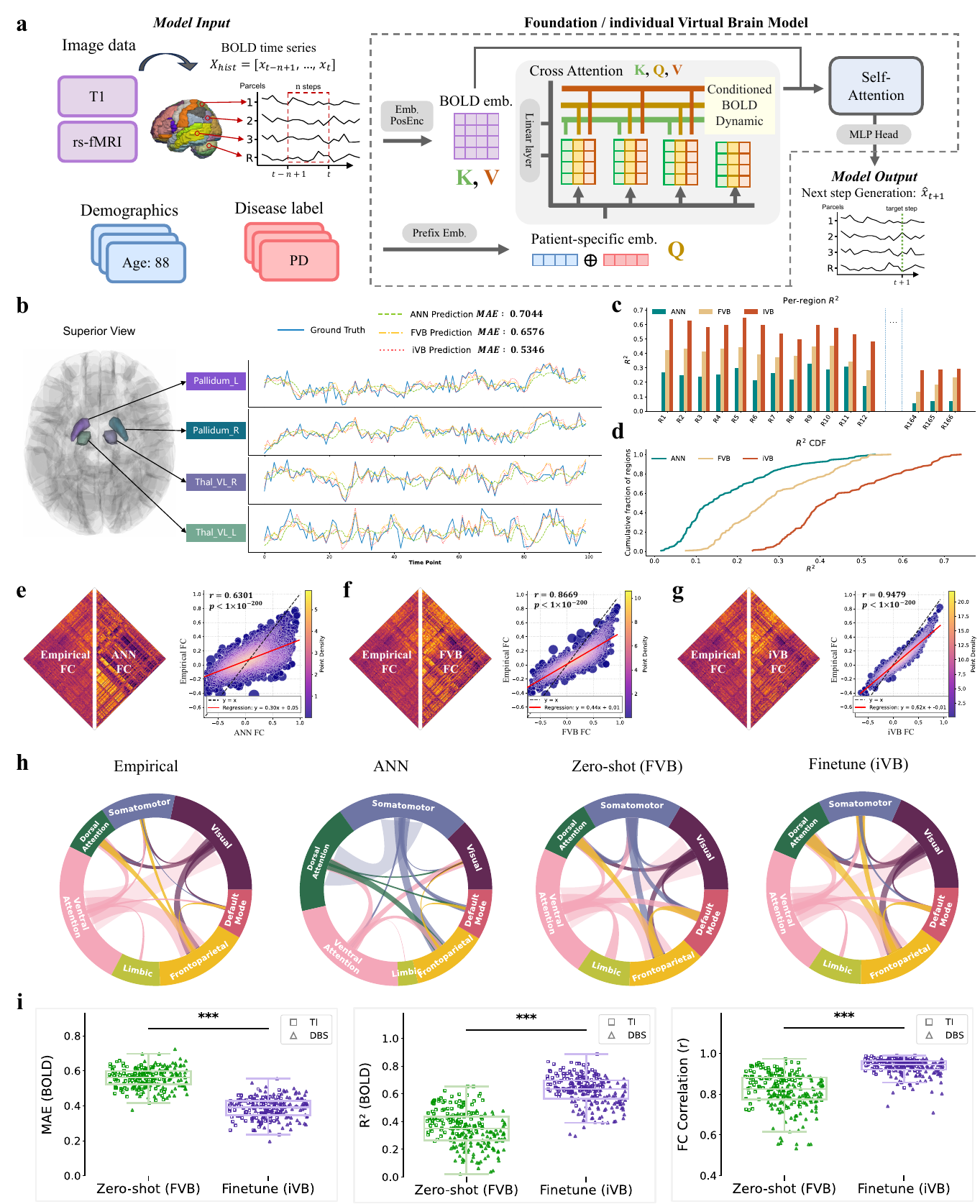}  
  \caption{
\textbf{Foundation and individualized virtual brain validation.}
\textbf{a,} Architecture of the generative virtual brain model.
\textbf{b,}
BOLD signal forecasting in four Parkinson's disease–relevant regions on the test set and global MAE across 166 regions: ANN = 0.7044, FVB = 0.6576, iVB = 0.5346.
\textbf{c,} Per-region R\textsuperscript{2} distribution across 166 brain areas.
\textbf{d,} Cumulative distribution function (CDF) of R\textsuperscript{2} values.
\textbf{e--g,} Functional connectivity validation for ANN (\textbf{e}; $r = 0.630$), FVB (\textbf{f}; $r = 0.867$), and iVB (\textbf{g}; $r = 0.948$) against empirical data (all $p < 10^{-200}$, two-tailed Pearson correlation).
\textbf{h,} 
Visualization of brain functional connectivity using Yeo 7-network parcellation illustrates that iVB preserves high-level connectivity patterns closely matching empirical data.
\textbf{i,} 
Boxplots \textit{w.r.t.} the performance comparison between FVB and iVB across 106 patients (TI and DBS cohorts) show the significant improvements in MAE, R\textsuperscript{2}, and FC correlation after finetuning (all $p < 0.001$, two-tailed paired t-test).
}
\vspace{-12pt}
  \label{fig:1}
\end{figure}

\textbf{The following results are organized to answer three sequential questions:} 
\begin{itemize}
    \item First, we evaluated whether the proposed foundation-to-individualized virtual brain pipeline can faithfully reproduce patient-specific brain dynamics, because credible counterfactual inference requires a biologically valid generative substrate. (See Fig.~\ref{fig:1}.)

    \item Second, we tested the central clinical question of this study: whether the iVB-based framework can predict individual response to TI and DBS better than existing neuroimaging and AI baselines, and whether its performance is sufficient to support potential clinical utility. (See Fig.~\ref{fig:3} and Fig.~\ref{fig:2.2.2}.)
    
    \item Third, we examined what biological insights these individualized models reveal, asking whether the same framework can nominate circuit-level and symptom-relevant signatures associated with treatment benefit, while also supporting interpretable explanations of model predictions. (See Fig.~\ref{fig:2.3.1}.)
\end{itemize}


\label{sec:2_result}

\subsection{High-fidelity brain dynamics modeling}
In this part, we systematically validated our generative virtual brain model (\figsubref{fig:1}{a}) at both the pretraining
level (FVB) and the individualized level (iVB) to demonstrate the ability to characterize brain dynamics, benchmarking against a recent state-of-the-art ANN model~\cite{luo2025mapping} across multiple complementary dimensions. 

\textbf{Improved regional BOLD forecasting accuracy.}
The FVB improved regional BOLD forecasting over the ANN baseline, and individualized finetuning further enhanced prediction accuracy, as evaluated on a held-out test set (10\% of pretraining cohort).
Specifically, a clear performance gradient emerged across models when evaluating representative predictions in selected PD-relevant regions of interest (e.g., the pallidum and ventrolateral thalamus) alongside whole-brain error statistics (\figsubref{fig:1}{b}).
Compared with the ANN baseline (MAE = 0.7044), the pretrained FVB substantially reduced forecasting error (MAE = 0.6576), and further finetuning yielded additional improvements in iVB (MAE = 0.5346).
Regional explanatory power was further quantified by R\textsuperscript{2} distributions across 166 brain regions (\figsubref{fig:1}{c}). The ANN model exhibited limited predictive fidelity across most regions, whereas the FVB showed widespread improvement. Finetuning to individual subjects further shifted the distribution toward higher R\textsuperscript{2} values, as confirmed by the cumulative distribution function analysis (\figsubref{fig:1}{d}). 
Together, these results indicate that pretraining yields the FVB that improves regional BOLD forecasting at the whole-brain level, while individualized finetuning further enhances subject-specific predictive fidelity. Notably, in PD-relevant basal ganglia-thalamic regions highlighted in \figsubref{fig:1}{b}, iVB most closely tracks the observed temporal dynamics, supporting its suitability for patient-level modeling of neuromodulation-relevant circuits.

\textbf{Better functional connectivity reconstruction.}
As accurate brain simulation also requires the preservation of large-scale network interactions, we similarly evaluated our virtual brain model on the functional connectivity (FC) reconstruction task.
In \figsubref{fig:1}{e--g}, we visualized the predicted FC matrices alongside the empirical FC matrix, and quantified correspondence using scatter–density comparisons of FC values.
\jc{As can be seen}, the ANN baseline showed only moderate correspondence with empirical FC patterns ($r = 0.630$), whereas the FVB substantially improved reconstruction fidelity ($r = 0.867$) (\figsubref{fig:1}{e–f}). After individual finetuning, iVB achieved a near-identical FC structure relative to empirical data ($r = 0.948$; \figsubref{fig:1}{g}; all $p < 10^{-200}$).
To summarize connectivity at the systems level, we further mapped the AAL3 parcellation used in the 166-region FC matrices onto the Yeo 7-network atlas~\cite{yeo2011organization}, enabling interpretation in terms of canonical functional systems (visual, somatomotor, dorsal attention, ventral attention, limbic, frontoparietal, and default mode networks) (\figsubref{fig:1}{h}).
This representation demonstrated that iVB preserved high-level inter-network organization closely matching empirical connectivity patterns.
In contrast, ANN predictions showed clear deviations, while the FVB captured global structure but failed to fully reproduce subject-specific connectivity features. 
These results indicate that FVB establishes a strong subject-generalizable prior for whole-brain connectivity. Building on this foundation, iVB yields a high-fidelity, near-empirical simulation of individual network topology, thereby providing a realistic substrate for subsequent counterfactual simulation.

\textbf{Consistent clinical cohort personalization.}
We next assessed generalization to real-world clinical data
by evaluating virtual brain performance in an in-hospital PD cohort from Ruijin Hospital (Shanghai), including
TI-treated patients ($n = 51$) and DBS-treated patients ($n = 55$).
For each patient, we rapidly finetuned the pretrained FVB via autoregressive prediction loss minimization on their pre-treatment resting-state fMRI data ($\leq 30$ epochs) to obtain an iVB.
Across the full clinical cohort, iVB showed consistently improved generative fidelity relative to the zero-shot FVB (\figsubref{fig:1}{i}): BOLD forecasting MAE decreased from 0.559 ± 0.056 (zero-shot FVB) to 0.385 ± 0.059 (iVB), R\textsuperscript{2} increased from 0.348 ± 0.128 to 0.625 ± 0.105, and FC correlation improved from r = 0.816 ± 0.087 to 0.935 ± 0.041 (all $p < 0.001$).
The distributional shifts across patients indicate that the improvements were broadly observed rather than driven by a small subset of individuals, demonstrating robust patient-level modeling fidelity on clinically acquired fMRI data.

\subsection{Validation of neuromodulation outcome prediction}
\begin{figure}[t!]
  \centering
  \includegraphics[width=0.98\linewidth]{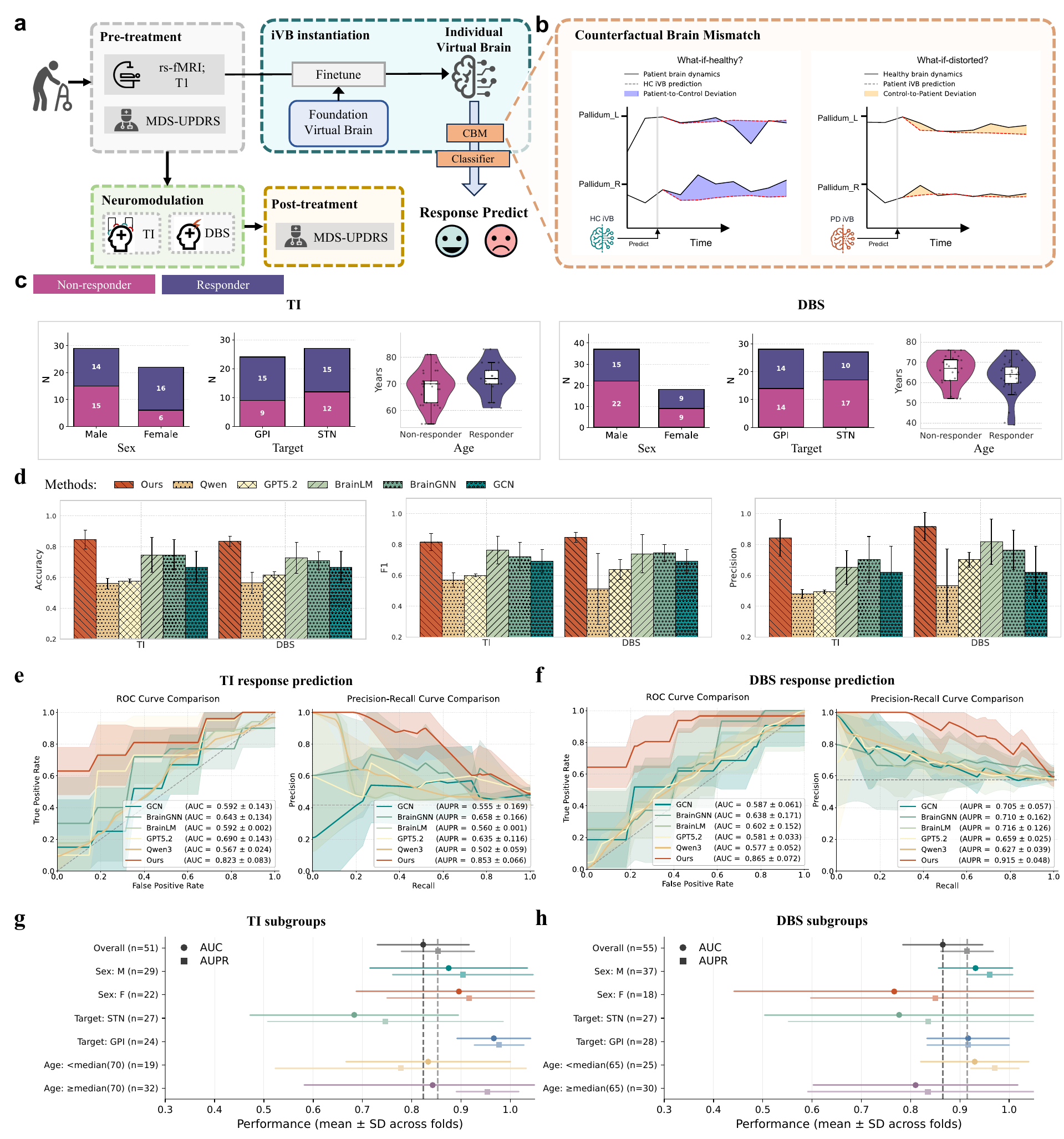}  
  \caption{\textbf{Validation of predicting neuromodulation response.}
  \textbf{a,} Schematic illustration of the iVB-based workflow for predicting neuromodulation response. 
  \textbf{b,} Diagram depicting the calculation of the Counterfactual Brain Mismatch.
  \textbf{c,} Clinical variable comparisons between responders/non-responders in TI ($n = 51$) and DBS ($n = 55$) cohorts. Fisher's exact test for sex (Male/Female) and target (GPi/STN); Mann–Whitney U test for age. All comparisons non-significant (ns; $p>0.05$).  
  \textbf{d--f,} Five-fold cross-validation performance comparing the proposed iVB-based model (Ours) with baselines. 
  \textbf{d,} Summary metrics (mean $\pm$ 95\% CI) for Accuracy (ACC), F1, and Precision. Ours achieved strong performance in both TI (ACC = 0.845, F1 = 0.816, Precision = 0.843) and DBS (ACC = 0.836, F1 = 0.847, Precision = 0.917), significantly outperforming all baselines (paired $t$-tests, all $p < 0.05$).
  Curves show mean ROC and PR with 95\% CI.
  \textbf{e,} TI cohort: AUC = $0.823 \pm 0.083$, AUPR = $0.853 \pm 0.066$ (ours). 
  \textbf{f,} DBS cohort: AUC = $0.865 \pm 0.072$, AUPR = $0.915 \pm 0.048$ (ours). 
  \textbf{g--h,} Forest plots report subgroup-wise AUC (circles) and AUPR (squares) across five cross-validation folds for the TI (\textbf{g}) and DBS (\textbf{h}).
  Logistic interaction tests showed no significant effect modification (all $p>0.05$)
}
  \label{fig:3}
\end{figure}

Building on the validated capacity of iVBs to capture individual brain dynamics, we deployed them within a prediction workflow (\figsubref{fig:3}{a}), which involves personalizing the FVB on individual neuroimaging data to obtain the iVB.
Central to this workflow is the extraction of Counterfactual Brain Mismatch (CBM) feature via bidirectional counterfactual simulations (\figsubref{fig:3}{b}), which quantified the divergence between patient-specific and normative brain dynamics in both What-if-healthy and What-if-distorted directions.
These CBM features serve as the decisive inputs for predicting neuromodulation outcomes, defined as $\geq 25\%$  improvement in MDS-UPDRS III scores for DBS and a reduction of $\geq 5$ points for TI. 

\textbf{Examining simple clinical covariates to remove confounding factors.} 
Here, we demonstrate that simple clinical covariates do not distinguish responders from non-responders in either the TI or DBS cohort.
As shown in \figsubref{fig:3}{c}, there were no significant differences between responders and non‑responders in sex (TI: $p = 0.19$; DBS: $p = 0.51$; Fisher’s exact test), stimulating target (GPi vs. STN; TI: $p = 0.19$; DBS: $p = 0.07$; Fisher’s exact test), or age (TI: $p = 0.33$; DBS: $p = 0.27$; Mann–Whitney U test).
These results indicate that treatment outcomes are not trivially separable by these conventional clinical variables. 
This absence of association reveals a critical clinical insight: neuromodulation outcomes in PD are not governed by traditional demographic or anatomical targeting variables. 
Instead, response heterogeneity likely arises from individual-specific pathophysiological network states invisible to conventional covariates.

\textbf{Comprehensive comparison in neuromodulation outcome prediction.}
Generally, the iVB-based framework significantly outperformed the established AI-driven baselines, achieving robust discrimination of responders versus non-responders across both therapy settings.
\jc{As shown in \figsubref{fig:3}{d--f}, we compared the prediction performance of our iVB-based method against several baselines: ChatGPT-5.2~\cite{singh2025openai} with deep-thinking few-shot learning, Qwen3~\cite{yang2025qwen3} with deep-thinking few-shot learning, BrainLM~\cite{caro2023brainlm} (a transformer-based fMRI model trained end-to-end), BrainGNN~\cite{li2021braingnn} (a graph neural network for fMRI classification), and GCN~\cite{kipf2016semi} (a discriminative graph convolutional network trained end-to-end on functional connectivity).}
In the TI cohort, our method achieved an ROC-AUC of 0.823 ± 0.083 and AUPR performance of 0.853 ± 0.066 (\figsubref{fig:3}{e}, showing mean curves with 95\% empirical confidence intervals [CI]). Consistent improvements were observed in the DBS cohort, where the model attained an AUC of 0.865 ± 0.072 and an AUPR of 0.915 ± 0.048 (\figsubref{fig:3}{f}).
Aggregate classification metrics further supported these improvements (\figsubref{fig:3}{d}). For TI, our approach achieved higher ACC (0.845 vs. best baseline BrainLM 0.747), F1 (0.816 vs. 0.765), and critically, Precision (0.843 vs. 0.703 by BrainGNN)---addressing the high cost of false positives in treatment trial-and-error.
For DBS, we similarly improved ACC (0.836 vs. 0.727), F1 (0.847 vs. 0.744), and Precision (0.917 vs. 0.817). All comparisons against baselines were statistically significant (all $p < 0.05$, paired t-tests).
To attribute these gains to our framework design, ablation analyses confirmed that CBM features substantially outperformed static FC, iVB-derived FC, or effective connectivity (Extended Data Table~\ref{tab:ablation_study}), validating the bidirectional counterfactual formulation.
Together, these results indicate that counterfactual mismatch quantified from patient-specific virtual brains provides clinically informative digital features for forecasting individual neuromodulation outcomes.
%
Our approach offers a novel and promising framework for the precision forecasting of individual neuromodulation.

\textbf{Stable robustness across different subgroups 
and effect modification.
}
To examine whether the predictive performance of our model varied across clinically relevant strata, we evaluated model discrimination within subgroups defined by sex (M/F), stimulation target (STN/GPi), and age (median split within each cohort).
As shown in \figsubref{fig:3}{g--h}, the model maintained robust performance across all subgroups in both TI and DBS cohorts, with subgroup-specific AUC and AUPR values comparable to overall performance.
We further tested whether subgroup membership modified the association between predicted probability and clinical response using logistic regression. No significant prediction–subgroup interactions were observed in either cohort (TI: Sex $p=0.161$, Target $p=0.092$, Age $p=0.534$; DBS: Sex $p=0.368$, Target $p=0.242$, Age $p=0.656$. All $p>0.05$). 
These findings suggest that the predictive utility of CBM is broadly preserved across conventional clinical strata, supporting the robustness of the framework across patient subgroups.

\begin{figure}[t!]
  \centering
  \includegraphics[width=0.95\linewidth]{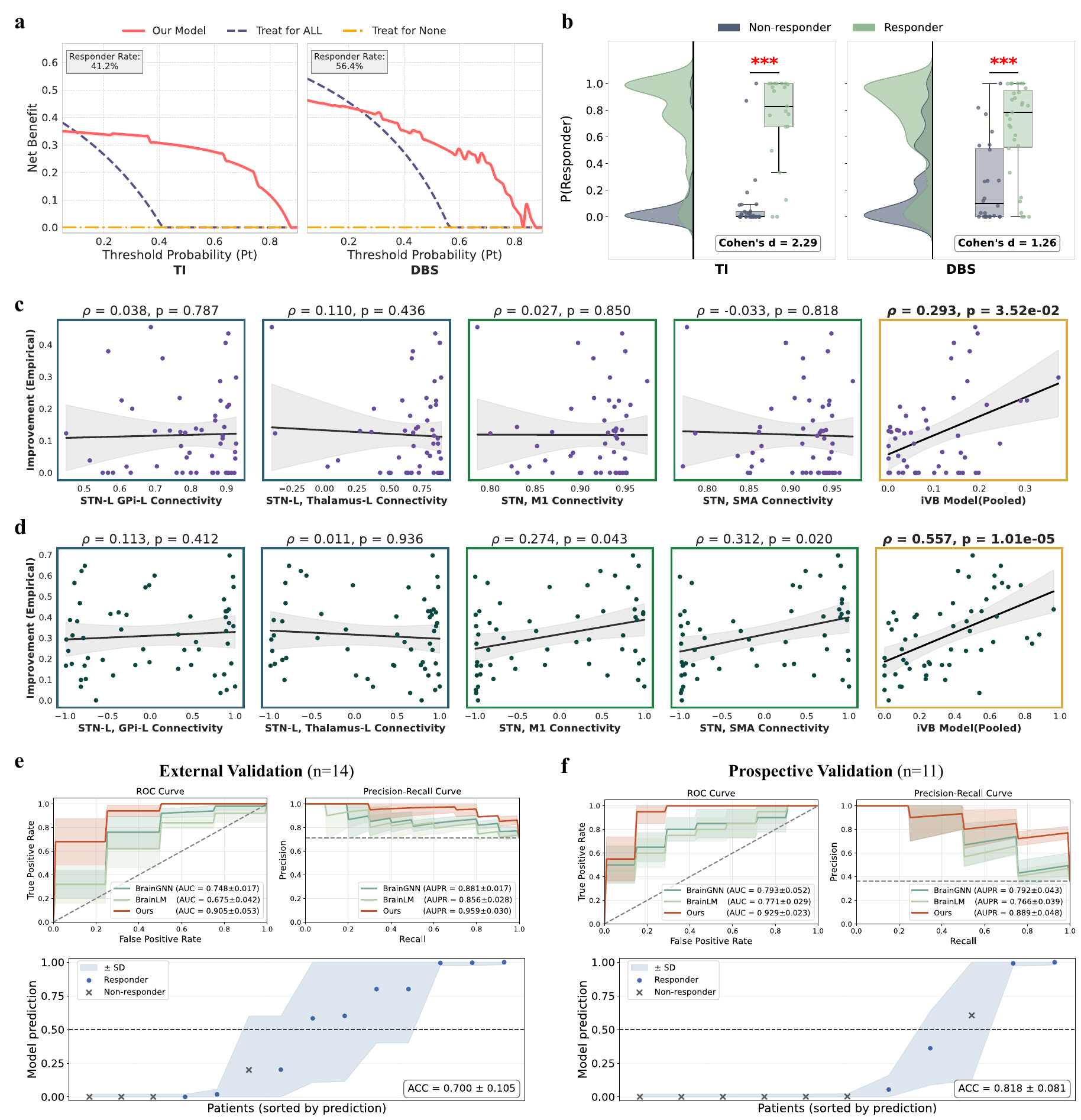}  
  \caption{\textbf{Clinical utility and validation of the iVB-based prediction framework.}
   \textbf{a,} Decision curve analysis (DCA) for the proposed model in TI and DBS cohorts. Our model (solid lines) outperforms reference strategies "Treat all" (dashed blue) and "Treat none" (dashed yellow). 
  \textbf{b,} Raincloud plots of predicted response probabilities stratified by clinical response status (Responders vs. Non-responders).
  Statistical comparisons: TI cohort (Mann-Whitney $U=33.00$, $p<0.001$, Cohen's $d=2.39$); DBS cohort (Mann-Whitney $U=132.00$, $p<0.001$, Cohen's $d=1.29$).
  \textbf{c--d,} Comparison of individual functional connectivity correlations versus integrated model prediction for TI (\textbf{c}) and DBS (\textbf{d}). The first four panels show Spearman correlations between clinical improvement and connectivity measures from different brain regions (left STN-GPi, left STN-Thalamus, STN-M1, STN-SMA). The fifth panel displays the iVB-based model results pooled from a five-fold cross-validation test data, which achieves significantly stronger correlation with treatment outcomes compared to individual connectivity measures.
  \textbf{e--f,} Model prediction ROC, and Precision-Recall curves, prediction spectrum plots, for external (\textbf{e}, n = 14) and prospective (\textbf{f}, n = 11) validation. Curves show mean ± std over five random seeds, with BrainLM and BrainGNN as baselines.
  }
  \label{fig:2.2.2}
\end{figure}

\textbf{Decision curve analysis to reflect clinical utility.}
Decision curve analysis (DCA) in \figsubref{fig:2.2.2}{a} demonstrates that the iVB-based framework yields a net clinical benefit across most threshold probabilities relevant to neuromodulation decisions.
Specifically, for both TI (responder rate 41.2\%) and DBS (56.4\%), the model provided a higher net benefit than the ``Treat for None'' strategies across nearly the entire range of thresholds.
Compared with the ``Treat for All'', net benefit was threshold-dependent: it was similar or lower at low Pt, but higher across most clinically relevant thresholds.
The advantage for TI emerged earlier and remained broadly sustained, while for DBS it was most pronounced in the mid-range of thresholds, consistent with DBS being a higher-burden intervention where avoiding over-estimating 
is especially important.
Overall, these results indicate meaningful clinical utility, improving responder targeting while reducing unnecessary procedures.

\textbf{Observed patient stratification by probabilistic confidence.} 
Complementing the threshold-based insights of DCA, the predicted response probability effectively served as a robust continuous biomarker to stratify responders from non-responders (\figsubref{fig:2.2.2}{b}).
The separation was particularly pronounced in the TI cohort (Mann-Whitney $U=33.00$, $p<0.001$, Cohen's $d=2.39$), with non-responders concentrated near probability 0 (<0.2) and responders near 1 (>0.8). The DBS cohort also maintained significant stratification (Mann-Whitney $U=132.00$, $p<0.001$, Cohen's $d=1.29$). 
Critically, prediction confidence itself directly functions as a clinical biomarker: probabilities >0.85 or <0.15 achieved 94\% specificity for true response status in TI, enabling risk-stratified patient selection. 
This probabilistic stratification transcends binary classification, offering clinicians interpretable risk gradients for personalized decision-making.

\textbf{Benchmarking against conventional neuroimaging biomarkers.}
The iVB-based predictor outperformed conventional single-pathway functional connectivity biomarkers using the identical pipeline
and continuous MDS-UPDRS III outcomes.
For this comparison, we established representative baselines reflecting standard clinical practice by selecting four key functional connectivity pathways previously reported to correlate with DBS treatment response: STN-L to GPi-L~\cite{younce2021resting}, STN-L to Thalamus-L~\cite{younce2021resting}, STN to M1~\cite{horn2017connectivity}, and STN to SMA~\cite{horn2017connectivity}. 
To accurately localize certain regions including STN, M1, and SMA, we adopted a combined labeling atlas incorporating cortical areas from Brainnetome~\cite{fan2016human} and subcortical areas from PD25~\cite{xiao2017dataset}. 
In the TI cohort (\figsubref{fig:2.2.2}{c}), individual connectivity measures showed no significant correlations with clinical improvement (all $p>0.4$). However, the iVB-based model achieved statistically significant correlation ($\rho = 0.293, p = 0.0352$), explaining approximately 8.6\% of variance in the response to treatment. In the DBS cohort (\figsubref{fig:2.2.2}{d}), individual connectivity pathways showed modest associations, with STN-M1 ($\rho = 0.274, p = 0.043$) and STN-SMA ($\rho = 0.312, p = 0.020$) connectivity reaching statistical significance. Notably, the iVB-based model substantially outperformed individual measures, demonstrating strong predictive capacity ($\rho = 0.557, p = 1.01\times10^{-5}$) and explaining approximately 31\% of variance in motor improvement outcomes.
These results demonstrate that while single-pathway biomarkers are susceptible to noise and insufficient to capture the complexity of neuromodulation effects, the high-dimensional approach using CBM effectively integrates distributed functional features to overcome inter-individual variability.

\textbf{Competitive performance under cross-center validation.}
To assess real-world generalizability, we performed an external validation on an independent TI cohort collected at Zhongnan Hospital of Wuhan University ($n = 14$). Responders were defined as patients showing a clinically meaningful benefit after repeated TI sessions (MDS-UPDRS III improvement $\geq 5$ points).
Without using any Zhongnan data for training, we applied the identical preprocessing and iVB construction pipeline, and directly transferred the TI response predictor trained on the Ruijin cohort to this external site.
We repeated inference over five random seeds and benchmarked against the top-performing prior baselines, BrainLM and BrainGNN.
Despite well-known cross-center heterogeneity in resting-state fMRI that can systematically degrade the portability of discriminative classifiers, our model retained strong performance on the held-out center (AUC = 0.905 ± 0.053, AUPR = 0.959 ± 0.030; \figsubref{fig:2.2.2}{e}), outperforming BrainGNN (AUC = 0.748 ± 0.017, AUPR = 0.881 ± 0.017) and BrainLM (AUC = 0.675 ± 0.042, AUPR = 0.856 ± 0.028).
Although the accuracy at the default 0.5 threshold was moderate (ACC = 0.700 ± 0.105), the prediction spectrum plot and strong AUC/AUPR indicate robust discriminative capacity despite cross-center heterogeneity.
We attribute this robustness to our generative, state-based formulation: instead of fitting a decision boundary on raw fMRI/FC features, the model leverages counterfactual simulations on iVBs to quantify deviations from normative brain dynamics, which is less sensitive to acquisition-related nuisance variation. Together, these results support the transferability of the iVB-based prediction model and the potential for multicenter deployment.

\textbf{Comparison in prospective clinical validation.}
We conducted a prospective validation study to evaluate the real-world clinical utility of our TI response prediction model.
The model, trained exclusively on retrospective data from Ruijin Hospital, was applied to consecutive, previously unseen Parkinson's disease patients ($n = 11$) scheduled for TI therapy.
Predictions were generated before treatment administration, and clinical outcomes were subsequently determined based on actual post-treatment assessments. 
As illustrated in \figsubref{fig:2.2.2}{f}, aggregating results over five random seeds, our model achieved an AUC of 0.929 ± 0.023 and AUPR of 0.889 ± 0.048, consistently surpassing BrainGNN (AUC = 0.793 ± 0.052, AUPR = 0.792 ± 0.043) and BrainLM (AUC = 0.771 ± 0.029, AUPR = 0.766 ± 0.039).
The model prediction spectrum also displays ranked probabilities for responders and non-responders, achieving an ACC of 0.818 ± 0.081.
This prospective demonstration provides critical evidence that virtual brain biomarkers can guide real-time therapeutic decisions in clinical workflows.
Although larger prospective trials are warranted, these initial results establish the feasibility of deploying individualized virtual brain models for precision neuromodulation planning.

\begin{figure}[!htbp]
  \centering
  \includegraphics[width=0.95\linewidth]{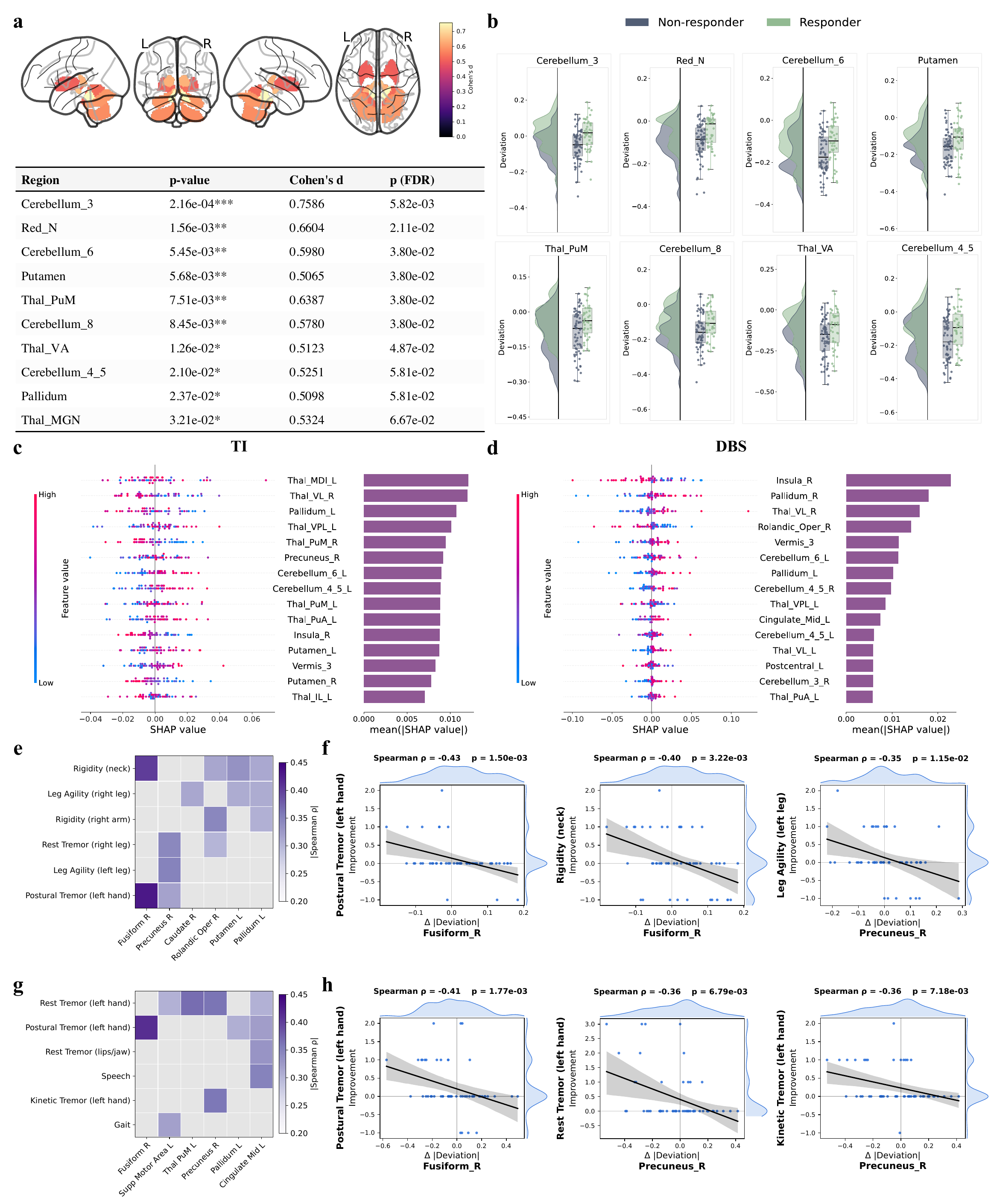}  
  \caption{ 
  \textbf{Mechanistic validation of iVB-based model via prediction, attribution, and symptom dynamics.}
  \textbf{a-b,}
  Pre-treatment iVB-derived regional CBM predicts TI response. 
  \textbf{a,} Brain map visualizes AAL3 regions most strongly associated with TI response, colored by Cohen’s $d$; inset table lists the regions meeting the dual threshold ($p < 0.05$, two-sided; $d > 0.5$), reporting uncorrected $p$, Cohen’s $d$, and FDR-adjusted $p$.
  \textbf{b,} Violin plots show CBM deviation distributions in these regions for responders versus non-responders.
  %
  %
  \textbf{c,} Shapley value analysis of the top 15 brain regions contributing to positive predictions in the TI cohort. \textbf{d,} Same Shapley value analysis for the DBS cohort.
  \textbf{e--h,} Region-symptom associations revealed by longitudinal iVB analysis.
  \textbf{e,} Heatmap showing Spearman correlations ($|\rho|$) between changes in regional deviation (post--pre) derived from iVBs and improvements in MDS-UPDRS III motor scores for TI cohorts. Only associations with $|\rho| \ge 0.3$ and $p < 0.05$ are displayed (non-significant values omitted).
  The panel shows a 6$\times$6 submatrix of the six most frequently implicated brain regions and MDS-UPDRS III items.
  \textbf{f,} Representative scatter plot illustrating the strongest region-symptom association with Spearman correlation coefficient and $p$-value indicated. 
  \textbf{g, h}, Same analysis as \textbf{e-f} for the DBS cohort, where \textbf{g} is the heatmap that characterizes the correlation and \textbf{h} gives the representative scatter plot. 
}
  \label{fig:2.3.1}
\end{figure}

\subsection{Mechanistic insights from the iVB-based model}
Beyond predictive accuracy, a clinically actionable virtual brain model should yield mechanistically interpretable insights that bridge computation and neurobiology.
Here, we interrogate our framework through three complementary lenses to establish its biological plausibility.
We first identify pre-treatment regional CBM that predict TI response, then deconstruct model decisions via Shapley attribution to reveal neuroanatomical drivers, and finally map longitudinal neurodynamic shifts onto symptom-specific improvements. 
All regional analyses utilize AAL3 brain labels (anatomical descriptions in Extended Data Table~\ref{tab:aal3_regions}).
%

\textbf{State-dependent digital biomarkers for TI response.}
%
As an emerging non-invasive neuromodulation, TI has been studied far less extensively than DBS, and the neural state dependencies underlying individual TI responses remain poorly characterized. 
Our iVB-based framework addresses this gap by quantifying patient-specific deviations from healthy brain dynamics and identifying regions where these deviations predict therapeutic outcomes.
We identified significant associations between pre-treatment regional activity mismatches and TI response (\figsubref{fig:2.3.1}{a}; statistics reported as p-values, Cohen's d effect sizes, and FDR-corrected q-values). 
Patients who subsequently responded to TI exhibited pronounced deviations from normative dynamics in a distributed network centered on the basal ganglia–thalamocortical circuit and cerebellar territories (\figsubref{fig:2.3.1}{b}). 
To balance discovery sensitivity with robustness against false positives, 
we employed a dual-threshold criterion (p < 0.05 and Cohen's d > 0.5).
Under this criterion, ten regions demonstrated substantial mismatch differences between responders and non-responders. The most prominent effects localized to cerebellar subregions (Cerebellum\_3: $d = 0.76$, $2.16 \times 10^{-4}$; Cerebellum\_6: $d = 0.60$), the red nucleus ($d = 0.66$), and canonical PD nodes including the putamen ($d = 0.51$) and pallidum ($d = 0.51$). 
Critically, these data-driven findings align with established circuit neuroanatomy and prior neuromodulation evidence: the pallidum constitutes a core component of the basal ganglia circuitry targeted by conventional DBS~\cite{deep2001deep}, with putaminal metabolic changes previously documented during successful neuromodulation~\cite{asanuma2006network}. 
The involvement of the ventral anterior thalamus further supports a plausible relay-level interpretation, given that VA/VL nuclei constitute canonical basal ganglia output relays with structured connections to frontal premotor territories~\cite{mcfarland2002thalamic}.
Moreover, the emergence of cerebellar lobules is consistent with tremor-relevant coupling between basal ganglia and the cerebello–thalamo–cortical loop, where GPi-related activity can propagate into cerebellar–thalamic circuitry~\cite{dirkx2016cerebral}.
The alignment of these data-driven findings with established neuroanatomy underscores the biological plausibility of the identified signatures and demonstrates the utility of virtual brain modeling for exploring mechanisms in precision neuromodulation.

\textbf{Model interpretability via Shapley analysis.}
Clinical deployment of AI demands transparency into how individual predictions are formed. 
Here, we leveraged Shapley value analysis~\cite{shapley1953value} to deconstruct the model's decision logic, 
which quantifies each brain region's marginal contribution to individual predictions, providing a mechanistically grounded interpretation of model decisions.
Specifically, we ranked regions by their mean absolute Shapley values and visualized the top 15 contributors to positive response predictions in \figsubref{fig:2.3.1}{c--d}.
Shapley attribution revealed distinct yet partially overlapping neuroanatomical drivers for the two therapies. For TI, the model's positive predictions were driven by a distributed pattern across fifteen brain regions with remarkably uniform influence magnitudes (mean absolute SHAP range: 0.0070–0.0121; max/min ratio $\approx$1.7).
Top contributors spanned thalamic relays (MDl, VL, VPL), cerebellar lobules (Crus VI, IV–V), and associative territories (precuneus, insula). 
This spatially diffuse signature aligns with TI's physical mechanism: interference of two high-frequency fields generates a focal envelope deep in the brain while simultaneously engaging off-target pathways through volume conduction~\cite{grossman2017noninvasive}.
Our model thus bases its decision on network-wide state shifts rather than isolated node modulation.
By contrast, Shapley values of DBS predictions exhibited a sharply concentrated contribution pattern (mean absolute SHAP range: 0.0058–0.0228; max/min ratio $\approx$3.9). 
The pallidum, VL thalamus, and peri-Rolandic cortices dominated the decision logic, with modest cerebellar contributions.
This focal architecture reflects DBS's localized current injection and its primary action on core nodes of the basal ganglia–thalamocortical motor circuit.
Notably, this demonstrates that the model internalized therapy-specific biophysical constraints without explicit programming.
Critically, by grounding prediction logic in established neuroanatomy and stimulation biophysics, Shapley analysis bridges data-driven forecasting and clinical mechanistic understanding, a prerequisite for trustworthy deployment of 
AI in neuromodulation planning.

\textbf{Longitudinal tracking reveals region-symptom mapping.}
Longitudinal changes in CBM-derived regional mismatch linked regional state normalization to symptom-specific motor improvement in both TI and DBS cohorts.
Using pre- and post-treatment fMRI scans, we constructed patient-specific virtual brains to quantify deviations from normative dynamics and track how therapeutic interventions reshaped pathological activity toward healthy states.
Strikingly, individual deviation trajectories within distinct neuroanatomical territories correlated with improvements in specific motor domains (\figsubref{fig:2.3.1}{e--h}, highlighting top associations; full matrices in Extended Data Fig.~\ref{fig:S1}), revealing a data-driven mapping between regional neurodynamics and symptom-specific recovery.
In the TI cohort (\figsubref{fig:2.3.1}{e--f}), symptom-linked brain regional changes were distributed across the basal ganglia and associative cortical territories. 
Notably, the mismatch change in pallidum/putamen showed consistent associations with improvements in rigidity and leg agility subitems, aligning with established striato-pallidal circuit involvement in akinetic-rigid motor features~\cite{mcgregor2019circuit, yu2013enhanced}. 
Beyond canonical motor nodes, cortical and higher-order regions (e.g., rolandic operculum and precuneus) also appeared among the most frequently implicated regions, suggesting that TI-related state shifts may engage broader integrative networks.
In the DBS cohort (\figsubref{fig:2.3.1}{g--h}), the mapping was more concentrated in midline control and thalamocortical nodes.
Changes in mid-cingulate co-varied with improvements in tremor-related sub-items and speech, consistent with the mid-cingulate contribution to cognitive aspects of motor control and action monitoring~\cite{hoffstaedter2014role}. 
Supplementary motor area changes tracked gait and kinetic tremor improvement, consistent with its established role as a core cortical hub for motor sequencing and gait control in PD~\cite{rahimpour2021supplementary}. 
Interestingly, an unexpected association was observed between the fusiform gyrus and postural tremor improvement in both the TI and DBS cohorts, suggesting a potential role for higher-order integrative regions in modulating tremor through state-dependent changes. 
Importantly, these relationships should be interpreted as exploratory and hypothesis-generating rather than causal evidence, given the modest sample sizes and the absence of preregistered symptom-level endpoints. 
Nevertheless, the emergence of anatomically and clinically plausible mappings supports the interpretability of iVB-derived state variables for interrogating treatment mechanisms.
Ultimately, these analyses illustrate that iVB-derived features provide interpretable state variables that can be mapped onto fine-grained symptom dimensions, enabling mechanistic hypothesis generation beyond binary outcome prediction.

\section{DISCUSSION}

\label{sec:3_discuss}
We present a generative, transferable virtual-brain framework that predicts individual neuromodulation outcome from resting-state fMRI while studying the potential mechanistic interpretation.
Specifically, we frame the outcome prediction as a dynamical-systems problem: rather than learning a cohort-specific decision boundary from high-dimensional fMRI features, we estimate each individual’s deviation from normative brain dynamics using the virtual brain models with the design of counterfactual simulation. 
This framework is well-suited to neuromodulation, where labeled cohorts are typically modest, treatment response is heterogeneous, and clinically useful models must remain portable and interpretable.

A central finding is that the proposed framework consistently outperformed existing AI and neuroimaging baselines in comprehensive comparisons spanning various performance metrics.
Across both TI and DBS, the iVB-based model outperformed all evaluated baselines on discrimination and, importantly, on precision-oriented metrics that are more relevant to treatment selection. This advantage was accompanied by robustness across subgroups, with no significant effect modification, and by ablation results showing that CBM features were more informative than static FC or alternative features.
Together, these findings suggest that separating transferable dynamical prior learning from subject-specific instantiation yields a more stable predictive framework than other modeling in neuromodulation datasets.

From a clinical perspective, the main value of the framework lies in improving responder identification under realistic threshold-based decision-making. 
In neuromodulation, false positives may lead to unnecessary invasive surgery (DBS) or prolonged ineffective treatment exposure (TI), whereas false negatives may delay beneficial therapy. 
The consistent gains in AUPR and precision observed here are therefore particularly important, as they reflect improved identification of true responders under realistic class imbalance and clinically conservative operating points. 
Moreover, decision curve analysis suggests that the model can deliver net benefit across clinically plausible thresholds, supporting a practical workflow in which predicted probability serves as a continuous stratification biomarker: patients with high predicted benefit can be prioritized, borderline cases can be routed to additional evaluation, and low-probability cases can be considered for alternative strategies or refined targeting rather than immediate escalation.

The framework also showed robust portability across sites and potential usability in real-world deployment settings. 
From a deployment perspective, a clinically useful predictor must remain actionable when applied to new patients and new acquisition settings, rather than requiring repeated site-specific retraining. 
In this study, we evaluated this practical requirement in two ways: external-center testing, where the TI predictor trained at Ruijin was applied to an independent cohort acquired at a different site without using any external-center data for model updating; 
and a small prospective setting, where predictions were generated before treatment in consecutive unseen patients and then compared against subsequently observed outcomes. 
Both evaluations provide preliminary evidence that the proposed pipeline can operate under real-world conditions with fixed preprocessing, iVB instantiation, CBM extraction, and probability output, without site-specific retraining.

Beyond prediction, the framework provides interpretable, region-resolved biomarkers rather than black-box predictions. 
Unlike models that rely on latent representations with limited clinical meaning, our framework produces region-resolved CBM features that are explicitly defined as counterfactual state deviations relative to a normative reference. 
This structure enables transparent auditability at the patient level: for a given prediction, clinicians can inspect which brain territories contribute most strongly to the estimated response probability (e.g., via Shapley attribution over CBM features), offering a circuit-level rationale.
%
Importantly, the same counterfactual iVB machinery is also hypothesis-generating. 
By localizing pre-treatment deviations associated with response and tracking post–pre CBM changes against symptom-level improvement, the model nominates state-dependent circuit signatures that can be studied in the future. 

Several limitations should be acknowledged. 
First, while transfer learning enhances data efficiency, the available labeled neuromodulation cohorts remain relatively small. This constraint is particularly pronounced for temporal interference (TI) therapy and prospective validation studies. The limited sample size reduces statistical power for meaningful subgroup analyses and introduces uncertainty when interpreting region-specific effects.
Second, iVBs are predominantly constructed from neuroimaging-derived signals (resting-state BOLD fMRI time series), which serve only as indirect proxies for underlying neural activity~\cite{logothetis2008we}. 
%
Although iVBs demonstrate high fidelity to empirical functional connectivity, they do not elucidate causal mechanisms of neuromodulation. 
Future efforts to bridge iVB dynamics with biophysical reality might integrate concurrent electrophysiological recordings, biophysical stimulation field models, and structural connectivity constraints.
Third, although the iVB formulation improves transparency (region-resolved deviations and Shapley-attributed contributors), the associations we report are statistical in nature and should not be interpreted as evidence of causal mechanisms. 
Given the dimensionality of regional features and modest cohort sizes, some associations may reflect chance correlations or residual confounding; moreover, multiple-comparison control (including FDR) reduces but does not eliminate false discoveries. 
Accordingly, these signatures are best viewed as hypothesis-generating leads that motivate preregistered validation in larger multi-center cohorts and targeted experimental tests (e.g., stimulation-parameter manipulations or multimodal corroboration).

The evidence collected from this study signals a practical convergence between generative, transferable virtual-brain modeling and precision neuromodulation decision-making in PD, demonstrating that the individualized virtual brain can support both accurate responder stratification and clinically interpretable circuit readouts across TI and DBS.
Looking ahead, a key opportunity is to move from prediction to actionable planning by using the generative framework to simulate candidate stimulation strategies (e.g., targets, waveforms, dosing) and prioritize those most likely to reduce pathological deviations toward normative dynamics.
Crucially, improving simulation realism will require iterative advances in the virtual-brain model itself. 
Future work could better bridge the model with biophysical reality by incorporating patient-specific structural connectivity (e.g., diffusion connectomes), biophysical stimulation field models (for DBS and TI), and concurrent electrophysiological recordings, enabling the neuromodulation effects to more faithfully capture stimulation propagation and support more reliable planning.
Realizing this potential will require larger preregistered multi-center prospective studies to establish calibration, portability, and decision impact, alongside comparative effectiveness evaluations against current selection practices. 
Together, these directions position the generative virtual brain model as a scalable path toward mechanism-guided, patient-specific neuromodulation beyond the trial-and-error manner.

\section{METHODS}

\label{sec:4_method}

\subsection{Participants}
\textbf{Study overview and ethical approval.} 
This study was conducted in accordance with the Declaration of Helsinki~\cite{world2013world}. 
For the TI cohort, the study was approved by the ethics committee of Ruijin Hospital, Shanghai Jiao Tong University School of Medicine (Approval No. 2025140), and this trial was registered at ClinicalTrials.gov (NCT06980935). 
For the external validation cohort, ethical approval was obtained from the Medical Ethics Committee of Zhongnan Hospital of Wuhan University (Approval No. 2026022). 
The DBS cohort is a retrospective study utilizing the same cohort used in the previous research~\cite{zhang2021subthalamic}. 
The study protocol was approved by the ethics committee of Ruijin Hospital, Shanghai Jiao Tong University School of Medicine (Approval No. 2018017). 
All participants provided written informed consent prior to data collection. Clinical data were de-identified to protect patient privacy.

\textbf{Recruitment and setting.} 
Participants were recruited from the Center of Excellence for Parkinson’s Disease, Ruijin Hospital, and the Department of Neurology at Zhongnan Hospital of Wuhan University.
The study comprised four distinct cohorts: 
Retrospective TI Cohort: Patients receiving Transcranial Interference (TI) therapy were recruited between May 2025 and December 2025 ($n = 51$);
Prospective TI Cohort: To validate real-world clinical utility, a consecutive series of patients was recruited starting January 2026 ($n=11$);
DBS Cohort: Patients undergoing Deep Brain Stimulation (DBS) were recruited between September 2019 and December 2020 ($n = 55$); 
External Validation Cohort: An independent TI cohort was collected at Zhongnan Hospital of Wuhan University between June 2025 and December 2025 ($n=14$);
Public datasets (ADNI, PPMI, ABIDE, HCP) were additionally used for foundation model pretraining but were not included in the clinical outcome analysis.

\textbf{Eligibility criteria.} 
Inclusion criteria for the TI Trial were: (1) diagnosis of idiopathic PD per Movement Disorder Society (MDS) criteria~\cite{postuma2015mds}; (2) Hoehn \& Yahr stage $\leq$ 3 in the on-medication state; and (3) stable dopaminergic medication ($\geq$4 weeks).
Exclusion criteria for the TI Trial included: (1) prior DBS or any other invasive therapy; (2) unstable cardiovascular, hepatic, or renal dysfunction; (3) active psychotic symptoms (e.g., hallucinations or delusions); or (4) personal or family history of epilepsy.
%
Inclusion criteria for the DBS trial were:
 (1) idiopathic PD per MDS criteria~\cite{postuma2015mds}; (2) clinical candidacy for DBS with bilateral quadripolar electrode implantation; (3) right-handedness.
%
Exclusion criteria comprised: (1) other serious psychiatric disorders meeting Diagnostic and Statistical Manual of Mental Disorders, Fifth Edition~\cite{edition2013diagnostic} (DSM-5) criteria; (2) other major neurological illnesses; (3) unstable vital signs; (4) excessive tremors in DBS-off/medication-off states; (5) any postoperative complications detected via postoperative MRI; (6) contraindications for MRI (e.g., non-MRI-conditional implants, except for DBS patients under specific safety protocols); (7) incomplete clinical follow-up data.

\textbf{Treatment protocols.} 
For the TI cohorts, stimulation was administered using NervioX-2400 targeting either the STN or GPi.
TI was delivered via scalp electrodes using 2 kHz and 2.13 kHz currents. Optimized electrode placement generated a 130-Hz interference field maximized at the STN or GPi.
Current intensities (1--4 mA per channel) were set to the maximum tolerable level while maintaining the simulation-derived ratio. Each session lasted 20 minutes.
For the DBS cohort, the patients underwent bilateral quadripolar DBS electrode implantation (Medtronic 3387, Medtronic, USA; or SceneRay 1210, SceneRay, China). 
Targets were either the GPi ($N = 28$) or the STN ($N = 27$). 
For scanning, DBS was turned off for at least one hour prior to imaging.
Clinical response was defined as a $\geq 25\%$ improvement in MDS-UPDRS III scores for DBS patients and a reduction of $\geq 5$ points for TI patients, assessed at post-treatment.

\subsection{Data overview}
\textbf{Public datasets for pretraining.}
The foundation virtual brain (FVB) model was pretrained using resting-state fMRI data from four large-scale public cohorts: 
ADNI (n = 219 participants, 697 sessions), PPMI (n = 579, 579), ABIDE (n = 1,097, 1,097), and the HCP (n = 812, 3,248), 
yielding a total of 5,621 fMRI sessions from 2,707 unique participants spanning Alzheimer’s disease (AD), mild cognitive impairment (MCI), Parkinson’s disease (PD), autism spectrum disorder (ASD), and healthy controls (HC). 
Comprehensive demographic and clinical characteristics are provided in Extended Data Table~\ref{tab:public_population}.
These heterogeneous datasets provide broad coverage of normative and pathological brain dynamics, enabling the FVB to learn subject-generalizable principles of whole-brain activity.

\textbf{Clinical cohort for finetuning and validation.} 
Clinical validation was performed using real-world Parkinson’s disease cohorts from Ruijin Hospital (Shanghai) and Zhongnan Hospital (Wuhan). 
All four cohorts (Retrospective TI, $n = 51$; Prospective TI, $n = 11$; DBS, $n = 55$; External TI Validation, $n = 14$) underwent pre-treatment resting-state fMRI and structural T1-weighted MRI acquisition for model input. 
Additionally, the retrospective TI and DBS cohorts included post-treatment fMRI scans, which were reserved for mechanistic analyses of treatment-induced brain dynamics but excluded from response prediction training to prevent data leakage. 
Longitudinal MDS-UPDRS III scores were collected for all patients to define clinical response. 
Detailed demographic and clinical characteristics are summarized in Extended Data Table~\ref{tab:clinical_population}.

\subsection{Neuroimaging preprocessing}

\textbf{Structural MRI preprocessing.} 
For volume-based cohorts (ADNI, PPMI, ABIDE, Clinical), T1-weighted images were skull-stripped using FMRIB Software Library (FSL)~\cite{jenkinson2012fsl}'s \texttt{bet} (Brain Extraction Tool). Tissue segmentation was performed with \texttt{fast} (FMRIB's Automated Segmentation Tool) to generate bias-corrected images. These were registered to MNI152 space via 12-DOF affine transformation (\texttt{flirt}, FMRIB's Linear Image Registration Tool) followed by nonlinear warping (\texttt{fnirt}, FMRIB's Nonlinear Image Registration Tool).
For the surface-based cohort (HCP), structural preprocessing was omitted as the fMRI data are stored in surface space.

\textbf{Resting-state fMRI preprocessing.} 
For volume-based cohorts, preprocessing included slice-timing correction (\texttt{slicetimer}), motion correction (\texttt{mcflirt}), and boundary-based EPI-to-T1 coregistration (\texttt{flirt} with normalized mutual information cost function).
Functional data were normalized to MNI152 space by combining EPI-to-T1 affine matrices with T1-to-MNI nonlinear warps (\texttt{applywarp}).
Nuisance regression was performed to remove confounding signals, including head motion, white matter, and cerebrospinal fluid signals. Volumes with excessive head motion (framewise displacement $> 0.5$ mm) were excluded.
For the HCP cohort, we utilized the preprocessed CIFTI data (grayordinate space), which inherently includes motion correction and surface registration.

\textbf{Time series extraction.} For volume-based cohorts: Mean BOLD signals were extracted from 166 AAL3 regions, yielding subject-specific time series matrices after detrending and bandpass filtering (0.01–0.1 Hz).
For the HCP cohort: Since data reside in the 91,282 grayordinate surface space (fs\_LR 32k mesh), we adapted the volumetric AAL3 atlas to this surface space to enable consistent extraction. 
This was achieved through a two-stage mapping procedure: (1) cortical AAL3 regions (MNI152 space) were resampled to the fs\_LR 32k surface via nearest-neighbor interpolation, with medial wall vertices masked using the HCP-MMP atlas; (2) subcortical AAL3 labels were aligned to HCP's subcortical mask and concatenated with cortical labels to form a 91,282-dimensional parcellation.
Mean signals were aggregated from the surface vertices within each of the AAL3 adapted regions, yielding time series matrices identical in dimension to the volume-based cohorts.

\textbf{Time series standardization.} To enhance model training stability and convergence, we applied per-region z-score normalization to all extracted time series. Specifically, for each brain region's time series across every session, we subtracted the mean and divided by the standard deviation (computed over timepoints). This transformation ensures zero-mean and unit-variance signals, mitigating regional amplitude disparities and improving numerical stability during model optimization.

\subsection{Foundation virtual brain structure and pretraining}
The FVB employs a transformer architecture designed to forecast whole-brain BOLD dynamics conditioned on clinical context.
For each subject, the full resting-state fMRI time series comprises $T_{\text{total}}$ timepoints after preprocessing. To eliminate hemodynamic initialization transients, the first 30 timepoints are discarded, yielding an effective sequence of length $T = T_{\text{total}} - 30$.
Demographic attributes (e.g., age, sex, race) are transformed into a fixed-length vector $\mathbf{d} \in \mathbb{R}^D$ via z-score normalization with mean imputation for continuous features, and one-hot encoding with an ``Unknown'' category for categorical features.
Formally, given a historical BOLD sequence $\mathbf{X}_{\text{hist}} = [\mathbf{x}_{t-n+1}, \dots, \mathbf{x}_t] \in \mathbb{R}^{n \times R}$ with $t\in\{n,\dots, T-1\}$ (where $R=166$ for AAL3), disease label $s$, and demographic vector $\mathbf{d}$, the model predicts $\mathbf{x}_{t+1} \in \mathbb{R}^R$.   

Each BOLD frame $\mathbf{X}_{\text{hist}}$ is first projected into a latent embedding space using a linear projection followed by positional encoding:
\begin{equation}
\mathbf{Z}_0 = \text{PosEnc}\big (\text{Linear} (\mathbf{X}_{\text{hist}})\big) \in \mathbb{R}^{n \times d_{\text{model}}},
\end{equation}
where $\text{Linear} (\cdot)$ is a learnable projection, and $\text{PosEnc} (\cdot)$ injects sinusoidal positional encodings. Disease context is encoded using patient-specific prefix embeddings $\mathbf{P}^s_{\text{dis}} \in \mathbb{R}^{d_{\text{model}}}$ inspired by distribution-characterizing representations~\cite{saito2023prefix}, while demographics $\mathbf{d}$ are mapped to $\mathbf{P}_{\text{dem}} = \text{MLP}_{\text{dem}} (\mathbf{d}) \in \mathbb{R}^{d_{\text{model}}}$. 
These embeddings are fused via element-wise addition:
\begin{equation}
\mathbf{P} = \mathbf{P}^s_{\text{dis}} + \mathbf{P}_{\text{dem}} \in \mathbb{R}^{d_{\text{model}}}.
\end{equation}
Clinical context modulates spatiotemporal dynamics through $J$ cross-attention layers. This formulation establishes a top-down modulation mechanism where static clinical priors actively retrieve spatiotemporal patterns from the BOLD sequence that are most salient to the patient's specific condition.
Specifically, the aggregation is defined as:
\begin{equation}
\begin{aligned}
    &\mathbf{c}_j = \text{CrossAttn}\left (\mathbf{Q}_j, \mathbf{K}_j, \mathbf{V}_j \right), \quad j = 1, \dots, J, \\
    &\text{with}\ 
    \begin{cases}
        \mathbf{Q}_j = \mathbf{c}_{j-1} \mathbf{W}^Q_j \\
        \mathbf{K}_j = \mathbf{Z}_0 \mathbf{W}^K_j, \\ \mathbf{V}_j = \mathbf{Z}_0 \mathbf{W}^V_j
    \end{cases}, \quad \text{and} \quad \mathbf{c}_0 = \mathbf{P},
\end{aligned}
\end{equation}
Here, $\mathbf{W}^K_j$, $\mathbf{W}^Q_j$, and $\mathbf{W}^V_j$ are learnable projection matrices for the key and value of the $j$-th layer.
The final output $\mathbf{c}_J \in \mathbb{R}^{1 \times d_{\text{model}}}$ represents the clinical-conditioned global context. 
To integrate this global prior back into the sequence, the context vector is expanded and fused via a residual connection:
\begin{equation}
\mathbf{I} = \mathbf{Z}_0 + \text{Expand} (\mathbf{c}_J),
\end{equation}
where $\text{Expand} (\cdot)$ broadcasts the context token to $n$ timepoints, injecting a global clinical prior into the BOLD sequence.
Subsequently, self-attention layers capture temporal dependencies:
\begin{equation}
\mathbf{H} = \text{SelfAttn} (\mathbf{I}) \in \mathbb{R}^{n \times d_{\text{model}}}.
\end{equation}
To derive a robust prediction state, we employ attention-based pooling to extract a global dynamic signature from the refined sequence:
\begin{equation}
\mathbf{h} = \sum_{\tau=1}^n \alpha_\tau \mathbf{H}_\tau, \quad \alpha_\tau = \frac{\exp (\mathbf{w}^\top \mathbf{H}_\tau)}{\sum_{k=1}^n \exp (\mathbf{w}^\top \mathbf{H}_k)},
\end{equation}
where $\mathbf{w} \in \mathbb{R}^{d_{\text{model}}}$ is a learnable weight vector. 
The forecasting head predicts $\mathbf{x}_{t+1}$:
\begin{equation}
\hat{\mathbf{x}}_{t+1} = \text{Linear}\big (\text{GELU} (\text{Linear} (\mathbf{h}))\big).
\end{equation}
The FVB was pretrained on all sessions from the ADNI, PPMI, ABIDE, and HCP datasets (totaling $N_{\text{sess}}$ sessions) with mean squared error (MSE) loss:
\begin{equation}
\mathcal{L}_{\text{pretrain}} = \frac{1}{N_{\text{sess}} \cdot T} \sum_{i=1}^{N_{\text{sess}}} \sum_{t=n}^{T-1} \|\mathbf{x}_{t+1}^{ (i)} - \hat{\mathbf{x}}_{t+1}^{ (i)}\|_2^2,
\end{equation}
using the AdamW~\cite{loshchilov2017decoupled} optimizer (learning rate $5 \times 10^{-5}$, weight decay $10^{-4}$) and a 9:1 train-validation split. Early stopping was applied when validation loss plateaued.

\subsection{Individualized virtual brain instantiation}
Patient-specific virtual brains were instantiated via rapid finetuning of the pretrained FVB. For a target patient $p$ with pre-treatment fMRI $\mathbf{X}_p$, disease label $s_p$, and demographics $\mathbf{d}_p$, we optimized the FVB parameters $\theta$ by minimizing:
\begin{equation}
\mathcal{L}_{\text{ft}} (\theta) = \frac{1}{T} \sum_{t=n}^{T-1} \|\mathbf{x}_{t+1} - f_\theta (\mathbf{x}_{t-n+1:t}, s_p, \mathbf{d}_p)\|_2^2,
\end{equation}
where $T$ is the sequence length.
Leveraging the robust spatiotemporal representations learned during pretraining, finetuning with a low learning rate ($10^{-6}$) enabled the model to quickly adapt to individual-specific dynamics.
This rapid convergence yields a personalized iVB that further enhances prediction accuracy for the target patient while avoiding overfitting through minimal parameter updates.

\subsection{Neuromodulation response prediction}
Our framework predicts individual response to TI/DBS by: (1) finetuning the FVB on pre-treatment fMRI to instantiate patient-specific iVBs, (2) computing CBM features through bidirectional virtual brain simulations, and (3) training a classifier that integrates CBM features with baseline clinical severity.

\textbf{CBM feature extraction.}
To characterize pathological brain dynamics, we introduce CBM, which computes a bidirectional mismatch derived from iVBs through counterfactual simulation. Our CBM captures clinically unobservable counterfactual scenarios: What would this patient's brain activity look like if it followed healthy dynamics? And how would this patient's brain distort normal activity patterns? By quantifying the mismatch between observed and counterfactual BOLD trajectories through bidirectional virtual brain simulations, we extract digital biomarkers that reveal hidden pathological states invisible to static imaging metrics.
Specifically, let $\mathcal{M}_{\text{norm}}$ denote a normative virtual brain (FVB conditioned on healthy controls; $s=\text{HC}$, $\mathbf{d}_{\text{mean}}^{\text{HC}}$), and let $\mathcal{M}_p$ denote the patient-specific iVB for subject $p$. For the preprocessed BOLD sequence $\mathbf{X}_p = [\mathbf{x}_{0}, \dots, \mathbf{x}_{T_{\text{total}}-1}] \in \mathbb{R}^{T \times R}$, we compute two directional mismatches:

\noindent{Patient-to-Control Deviation}: How much the patient's dynamics deviate from normative expectations:  
\begin{equation}
\delta_p^{\text{ (PtC)}} = \frac{1}{T} \sum_{t=n}^{T-1} \big\| \mathbf{x}_{t+1} - \mathcal{M}_{\text{norm}} (\mathbf{x}_{t-n+1:t}, s=\text{HC}, \mathbf{d}_{\text{mean}}^{\text{HC}}) \big\|_2 \in \mathbb{R}^R
\end{equation}
\noindent{Control-to-Patient Deviation}: How the patient's brain model distorts healthy activity patterns. For $K$ representative healthy subjects ($\mathbf{X}_{\text{HC}}^{ (k)}$):
\begin{equation}
\delta_p^{\text{ (CtP)}} = \frac{1}{K} \sum_{k=1}^K \frac{1}{T} \sum_{t=n}^{T-1} \big\| \mathbf{x}_{t+1}^{\text{ (HC},k)} - \mathcal{M}_p (\mathbf{x}_{t-n+1:t}^{\text{ (HC},k)}, s_p, \mathbf{d}_p) \big\|_2 \in \mathbb{R}^R
\end{equation}
The regional CBM feature vector concatenates these directional mismatches per brain region:
\begin{equation}
\boldsymbol{\phi}_p^{\text{CBM}} = \big[ \delta_p^{\text{ (PtC)}}, \delta_p^{\text{ (CtP)}} \big] \in \mathbb{R}^{2R}
\end{equation}

\textbf{Response prediction.}
CBM features are combined with clinical variables to predict treatment response. For patient $p$, the input feature vector integrates regional CBM deviations and MDS-UPDRS III score:
\begin{equation}
\label{eq:zp}
\mathbf{z}_p = \big[ \boldsymbol{\phi}_p^{\text{CBM}}, \mathbf{u}_p^{\text{pre}} \big] \in \mathbb{R}^{2R + u},
\end{equation}
where $ \mathbf{u}_p^{\text{pre}} \in \mathbb{R}^{u}$ denotes the pre-treatment MDS-UPDRS III feature vector comprising $u$ individual item scores or subscores. A three-layer MLP with hidden dimensions 512 and 256 maps $\mathbf{z}_p$ to a response probability:
\begin{equation}
\label{eq:arch}
\hat{y}_p = \sigma\big (f_{\text{pre}} (\mathbf{z}_p)\big),
\end{equation}
where  $f_{\text{pre}}$  denotes the MLP function and $\sigma$ is the sigmoid activation. The model is trained via binary cross-entropy loss:
\begin{equation}
\mathcal{L}_{\text{cls}} = -\frac{1}{M} \sum_{m=1}^M \Big[ y_m \log (\hat{y}_m) + (1-y_m)\log (1-\hat{y}_m) \Big],
\end{equation}
where $y_m=1$ denotes responders ($\geq 25\%$  improvement in MDS-UPDRS III scores for DBS; a reduction of $\geq 5$ points for TI). Five-fold cross-validation used stratified splits preserving responder ratios.

\textbf{Improvement rate regression.}
Beyond binary response classification, we performed regression analysis to predict continuous improvement rates. For patient $p$, the empirical improvement rate is defined as:

\begin{equation}
I_p = \max\Big\{\frac{\sum (\mathbf{u}_p^{\text{pre}} - \mathbf{u}_p^{\text{post}})}{\sum\mathbf{u}_p^{\text{pre}}},0\Big\},
\end{equation}
where $ \mathbf{u}_p^{\text{pre}}$ and $\mathbf{u}_p^{\text{post}}$ denote the pre- and post-treatment MDS-UPDRS III feature vector, respectively.

For patient $p$, we construct the input feature vector \(\mathbf{z}_p\). The predicted rate is $\hat{I}_p = \hat{y}_p = \sigma (f_{\text{pre}} (\mathbf{z}_p))$ as Equation~\ref{eq:arch}.
The model is trained by minimizing a hybrid loss function that combines mean squared error with a differentiable ranking objective:

\begin{equation}
\mathcal{L}_{\text{reg}} = \alpha \mathcal{L}_{\text{MSE}} + \beta \mathcal{L}_{\text{rank}},
\end{equation}

where $\alpha$ and $\beta$ are weighting hyperparameters. The MSE component ensures accurate numerical prediction:

\begin{equation}
\mathcal{L}_{\text{MSE}} = \frac{1}{N} \sum_{n=1}^N \big ( I_n - \hat{I}_n \big)^2,
\end{equation}

while the ranking component $\mathcal{L}_{\text{rank}}$ maximizes the Spearman correlation between predicted and observed improvement rates. To enable gradient-based optimization, we employ a soft-ranking approximation using temperature-scaled sigmoid functions. For a batch of predictions $\{\hat{I}_n\}_{n=1}^N$, the soft rank of each sample is computed as:

\begin{equation}
r (\hat{I}_i) = 1 + \sum_{j=1}^N \sigma\left (\frac{\hat{I}_i - \hat{I}_j}{\tau}\right),
\end{equation}

where $\sigma (\cdot)$ is the sigmoid activation and $\tau$ is the temperature parameter. The ranking loss is then defined as:

\begin{equation}
\mathcal{L}_{\text{rank}} = 1 - \rho_s\big (r (\hat{\mathbf{I}}), r (\mathbf{I})\big),
\end{equation}

where $\rho_s$ denotes the Spearman correlation coefficient computed from the soft ranks. This formulation encourages the model to preserve the relative ordering of improvement rates, which is particularly valuable given the ordinal nature of clinical outcome measures and the potential presence of outliers.

We employed stratified five-fold cross-validation, where patients were partitioned into five folds while preserving the distribution of improvement rates. The model was trained on four folds and evaluated on the held-out fold, repeating this for each fold. Final pooled performance metrics (Spearman correlation R and p-value) were computed by collecting predictions from all five test sets and calculating the correlation between the predicted and empirical improvement rates, yielding a fair comparison between other non-training biomarkers.

\subsection{Performance metrics}
\textbf{Generative virtual brain fidelity.}
We evaluated the fidelity of the foundation virtual brain (FVB) and individualized virtual brain (iVB) in forecasting regional BOLD dynamics and reconstructing functional connectivity (FC).
Given ground-truth next-step BOLD vector $\mathbf{x}_{t+1}\in\mathbb{R}^{R}$ and the model prediction $\hat{\mathbf{x}}_{t+1}$, forecasting error was quantified by mean absolute error (MAE):
\begin{equation}
\mathrm{MAE}=\frac{1}{T}\sum_{t=1}^{T}\frac{1}{R}\sum_{r=1}^{R}\left|x_{t,r}-\hat{x}_{t,r}\right|.
\end{equation}
We additionally reported the coefficient of determination ($R^{2}$), computed per region and then averaged across regions:
\begin{equation}
R^{2}=1-\frac{\sum_{t} (x_{t,r}-\hat{x}_{t,r})^{2}}{\sum_{t} (x_{t,r}-\bar{x}_{r})^{2}},\quad \bar{x}_{r}=\frac{1}{T}\sum_{t}x_{t,r}.
\end{equation}
For FC reconstruction, we computed Pearson correlation between vectorized upper-triangular entries of the empirical FC matrix $\mathbf{C}$ and the predicted FC matrix $\hat{\mathbf{C}}$:
\begin{equation}
r_{\mathrm{FC}}=\mathrm{corr}\left (\mathrm{vec} (\mathbf{C}_{\triangle}),\,\mathrm{vec} (\hat{\mathbf{C}}_{\triangle})\right),
\end{equation}
where $\triangle$ denotes the upper triangle excluding the diagonal.

\textbf{Neuromodulation response prediction.}
For binary responder classification, we assessed discrimination using the area under the receiver operating characteristic curve (ROC-AUC) and the area under the precision--recall curve (AUPR), which is particularly informative under class imbalance.
Threshold-dependent metrics included Accuracy, Precision, and F1-score.
Confusion matrices were reported in normalized form.
To further quantify decision impact, we performed decision curve analysis (DCA) and computed net benefit (NB) across threshold probabilities $p_t$:
\begin{equation}
\mathrm{NB} (p_t)=\frac{TP}{N}-\frac{FP}{N}\cdot\frac{p_t}{1-p_t},
\end{equation}
where $N$ is the cohort size. Reference strategies ``treat all'' and ``treat none'' were included as baselines.
When evaluating agreement between predicted and observed continuous clinical improvement, we reported Spearman rank correlation ($\rho$) between predicted improvement and measured MDS-UPDRS III change, along with two-sided $p$-values.


\subsection{Statistical analysis}
All statistical tests were two-sided with a significance level of 0.05. 
We report summary statistics as mean $\pm$ standard deviation (SD) or with 95\% confidence intervals (CI).
For comparisons of continuous variables, we assessed normality using the Shapiro--Wilk test.
When normality was not violated, we used parametric tests (independent or paired $t$-tests as appropriate).
Otherwise, we used non-parametric alternatives (Mann--Whitney U test for independent samples and Wilcoxon signed-rank test for paired samples).
Categorical variables were compared using Fisher's exact test or $\chi^2$ test.
Subgroup performance was summarized across five cross-validation folds (mean $\pm$ SD), and effect modification was tested using logistic regression with prediction-subgroup interaction terms.
Effect sizes for group differences were quantified using Cohen's $d$ for continuous variables. For region-wise analyses across brain parcellations, we controlled the false discovery rate (FDR) using the Benjamini--Hochberg procedure.

\subsection{Computational hardware and software}
All MRI and fMRI data were processed on a high-performance workstation equipped with an Intel Xeon Gold 6530 processor (128-core, 2.1 GHz) and 8 NVIDIA GeForce RTX 4090 GPUs. Our software development utilized Python (version 3.9.20) and the models were developed using PyTorch (version 2.1.2) and transformers (version 4.38.2). We used several other Python libraries to support data analysis and neuroimaging, including MONAI (version 1.3.0), nibabel (version 5.3.2), nilearn (version 0.12.0), pandas (version 2.2.3), scipy (version 1.13.1), and scikit-learn (version 1.6.0). MRI and fMRI preprocessing was conducted using dcm2niix and FSL (version 6.0.7.18). Training the FVB model on a single RTX 4090 GPU during the pre-training phase had an average runtime of 3 minutes per epoch, whereas the iVB finetuning task took less than 30 seconds per epoch. Figures were prepared using Matplotlib and Seaborn.
\section{Data availability}
The ADNI, ABIDE, and PPMI datasets are publicly available through the Image \& Data Archive (IDA) hosted by the Laboratory of Neuro Imaging (LONI) at the University of Southern California (\url{https://ida.loni.usc.edu}). 
The Human Connectome Project (HCP) Young Adult 1200 Subjects data release is accessible via the HCP portal (\url{https://www.humanconnectome.org/study/hcp-young-adult/document/1200-subjects-data-release}).  
The AAL3 brain atlas employed in this work is publicly accessible via \url{https://www.gin.cnrs.fr/en/tools/aal/}.
The PD25 atlas is available through the NeuroImaging\&Surgical Technologies Lab (NIST) at  \url{https://nist.mni.mcgill.ca/multi-contrast-pd25-atlas/}.
The Brainnetome atlas can be accessed at \url{https://atlas.brainnetome.org/}.

The fMRI data from the Ruijin Hospital and Zhongnan Hospital are not publicly available due to institutional data privacy regulations governing patient information. 
However, the study protocol, de-identified preprocessed fMRI derivatives, and associated clinical assessments are available from the corresponding author upon reasonable request, subject to ethical approval and execution of a data sharing agreement, to support reproducibility and independent verification.

\section{Code availability}
The code supporting this study is available on GitHub at \url{https://github.com/desomeboy/Foundation-Individual-Generative-Virtual-Brain}, under the Apache License, v.2.0 (Apache-2.0). 

\section{Acknowledgements}
This work is supported by the National Key R\&D Program of China (No. 2022ZD0160702),  National Natural Science Foundation of China (No.62306178) and STCSM (No. 22DZ2229005), 111 plan (No. BP0719010).

\section{Author contributions}
S.D. and S.L. conceived the idea, curated public data, implemented code, developed and validated the framework, visualized results and wrote the manuscript. S.B., A.L. and Y.P. collected clinical data and administered treatment (S.B. assisted in mechanism writing; A.L. preprocessed clinical data). H.L. assisted in visualization and writing. M.X. designed the framework and guided experimental design. D.L. guided experimental design and revised the manuscript. W.X., Y.W., Y.Z., C.Z. and J.Y. guided experimental design, wrote and revised the manuscript. 

\section{Competing interests}
The authors declare no competing interests.
\newpage
\bibliographystyle{unsrt} 
\bibliography{references} 

\clearpage
\appendix
\setcounter{figure}{0}
\renewcommand{\figurename}{Extended Data Fig.}

\begin{figure}[ht]
  \centering
  \includegraphics[width=0.98\linewidth]{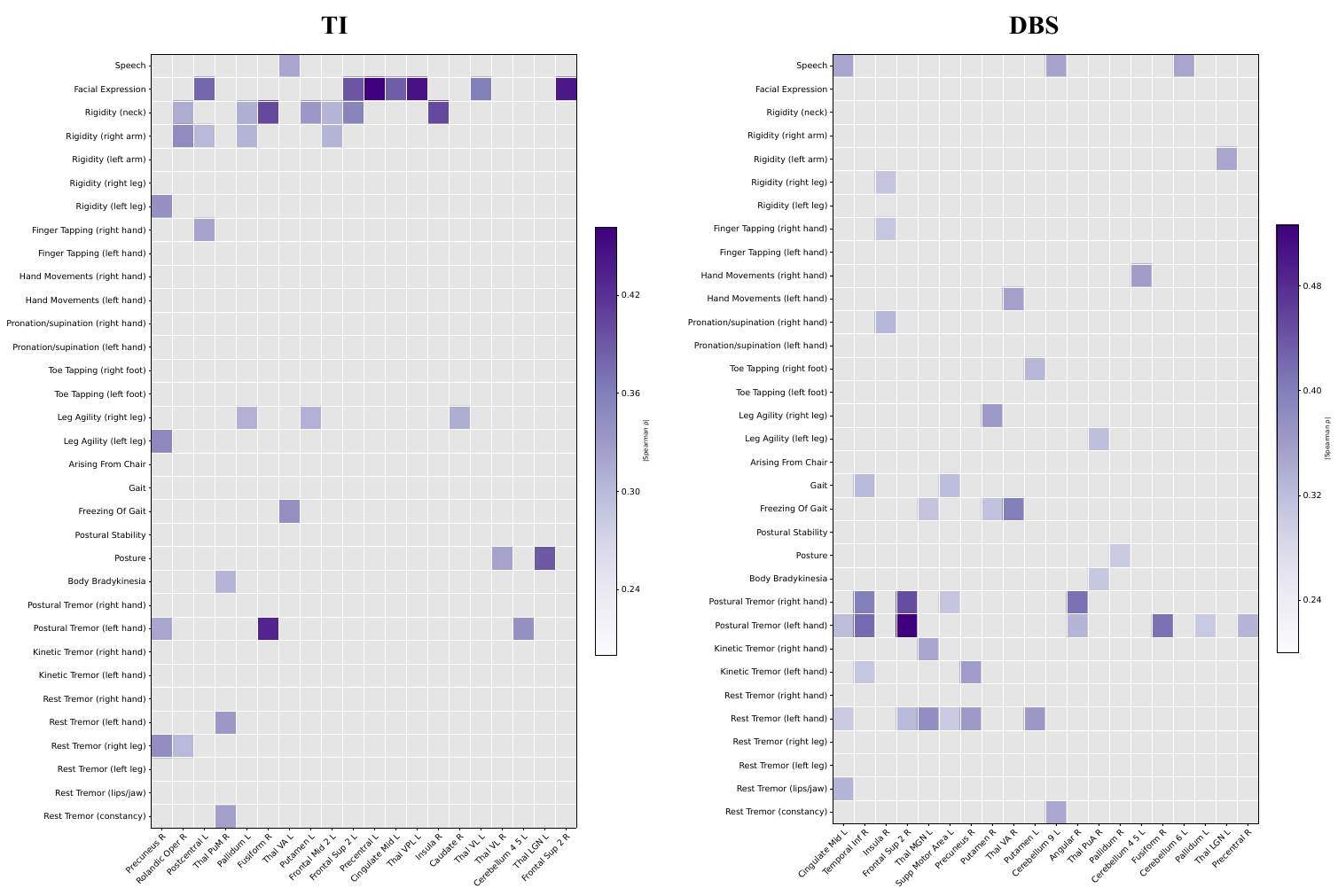}
  \caption{Full Heatmap of absolute Spearman correlation strengths between brain regions and MDS-UPDRS symptom measures. Rows denote individual symptoms. Columns show the top-ranked regions, defined by the number of associations exceeding the threshold ($|\rho| \ge 0.3$ and $p < 0.05 $). Color represents $|\rho|$, with darker shades indicating stronger effects. Associations not meeting the threshold or not confirmed by the significance table are displayed in gray.
  Results are presented for TI and DBS cohorts separately.}
  \label{fig:S1}
\end{figure}


\begin{table}[h!]
    \centering
    \captionsetup{name=Extended Data Table}
    \caption{\textbf{Ablation Study on CBM}}
    \label{tab:ablation_study}
    \setlength{\tabcolsep}{8pt} 
    \setlength{\extrarowheight}{6pt}
    \small
    \begin{tabular}{m{2.5cm} m{2.2cm} m{2.2cm} m{2.2cm} m{2.2cm} m{2.2cm}}
    \hline
    \rowcolor[gray]{0.9}
    \textbf{Feature Type} & \textbf{AUC} & \textbf{AUPR} & \textbf{ACC} & \textbf{F1} & \textbf{Precision} \\
    \hline
    Static\_FC & 0.417 ± 0.283 & 0.609 ± 0.167 & 0.600 ± 0.178 & 0.418 ± 0.378 & 0.450 ± 0.371 \\
    \hline
    iVB\_FC & 0.420 ± 0.152 & 0.578 ± 0.115 & 0.582 ± 0.136 & 0.544 ± 0.276 & 0.568 ± 0.326 \\
    \hline
    iVB\_EC & 0.784 ± 0.090 & 0.850 ± 0.076 & 0.727 ± 0.100 & 0.773 ± 0.074 & 0.781 ± 0.158 \\
    \hline
    \rowcolor[gray]{0.95}
    CBM & \textbf{0.865 ± 0.072} & \textbf{0.915 ± 0.048} & \textbf{0.836 ± 0.036} & \textbf{0.847 ± 0.036} & \textbf{0.917 ± 0.105} \\
    \hline
    \end{tabular}
    \begin{flushleft}
    {\small Ablation study on Counterfactual Brain Mismatch (CBM) features for neuromodulation response prediction. Four strategies evaluated via stratified 5-fold CV on retrospective DBS cohorts; all features concatenated with pre-treatment MDS-UPDRS III scores.
    Static\_FC: Pre-treatment resting-state fMRI FC. iVB\_FC: FC from iVB-generated BOLD.  
    iVB\_EC: Effective connectivity map via perturbing the iVB surrogate brain to estimate directed region-to-region influences, following the protocol in Luo et al.~\cite{luo2025mapping}.
    CBM: bidirectional counterfactual brain mismatch. 
    Values: Mean ± SD; best in bold.}
    \end{flushleft}
\end{table}

\begin{table}[h]
\centering
\captionsetup{name=Extended Data Table}
\caption{\textbf{AAL3 Atlas Region Labels in Main Text and Corresponding Anatomical Names.} }
\label{tab:aal3_regions}
\setlength{\extrarowheight}{8pt}
\resizebox{\textwidth}{!}{%
\begin{tabular}{ll|ll}
\hline
\rowcolor[gray]{0.9}
\textbf{AAL3 Label} & \textbf{Anatomical Name} & \textbf{AAL3 Label} & \textbf{Anatomical Name} \\[2ex]
\hline
Angular & Angular Gyrus & Putamen & Putamen \\
Caudate & Caudate Nucleus & Red\_N & Red Nucleus \\
Cerebellum\_3 & Lobule III of Cerebellar Hemisphere & Rolandic\_Oper & Rolandic Operculum \\
Cerebellum\_4\_5 & Lobule IV, V of Cerebellar Hemisphere & Supp\_Motor\_Area & Supplementary Motor Area \\
Cerebellum\_6 & Lobule VI of Cerebellar Hemisphere & Temporal\_Inf & Inferior Temporal Gyrus \\
Cerebellum\_8 & Lobule VIII of Cerebellar Hemisphere & Thal\_IL & Intralaminar Thalamic Nucleus \\
Cerebellum\_9 & Lobule IX of Cerebellar Hemisphere & Thal\_LGN & Lateral Geniculate Nucleus \\
Cingulate\_Mid & Middle Cingulate \& Paracingulate Gyri & Thal\_MDl & Mediodorsal Lateral Thalamic Nucleus \\
Frontal\_Mid\_2 & Middle Frontal Gyrus & Thal\_MGN & Medial Geniculate Nucleus \\
Frontal\_Sup\_2 & Superior Frontal Gyrus, Dorsolateral & Thal\_PuA & Pulvinar Anterior Thalamic Nucleus \\
Fusiform & Fusiform Gyrus & Thal\_PuM & Pulvinar Medial Thalamic Nucleus \\
Insula & Insula & Thal\_VA & Ventral Anterior Thalamic Nucleus \\
Pallidum & Globus Pallidus & Thal\_VL & Ventral Lateral Thalamic Nucleus \\
Postcentral & Postcentral Gyrus & Thal\_VPL & Ventral Posterolateral Thalamic Nucleus \\
Precentral & Precentral Gyrus & Vermis\_3 & Lobule III of Vermis \\
Precuneus & Precuneus & & \\
\hline
\end{tabular}
}
\begin{flushleft}
    {\small Note that each AAL3 label represents a brain region, with suffixes \_L and \_R denoting the left and right hemisphere, respectively.}
    \end{flushleft}
\end{table}

\begin{table}[h!]
    \centering
    \captionsetup{name=Extended Data Table}
    \caption{\textbf{Public Dataset Study Population} }
    \label{tab:public_population}
    \setlength{\tabcolsep}{12pt}
    \setlength{\extrarowheight}{8pt}
    \small
    \begin{tabular}{m{1.2cm} m{0.6cm} m{1.6cm} m{2.1cm} m{2.8cm} m{3.5cm}}
    \hline
    \rowcolor[gray]{0.9}
    \textbf{Dataset} & \textbf{n} & \textbf{Age (y)} & \textbf{Male n (\%)} & 
    \textbf{\makecell{Race \\(W, B, A, I, H, O)}} & \textbf{Label Distribution} \\
    \hline
    HCP & 812 & N/A & 402 (49.5) & 887, 193, 64, 2, 5, 45 & \makecell[l]{HC: 812} \\
    \hline
    PPMI & 579 & 65.9 ± 13.9 & 326 (56.3) & 559, 4, 7, 3, 0, 6 & 
    \makecell[l]{HC: 406; PD: 173} \\
    \hline
    ABIDE & 1,097 & 17.0 ± 8.0 & 934 (85.1) & N/A & 
    \makecell[l]{HC: 570; \\Autism: 527} \\
    \hline
    ADNI & 219 & 73.1 ± 7.2 & 101 (46.1) & 197, 11, 4, 0, 0, 7 & 
    \makecell[l]{HC: 82; MCI: 103; \\AD: 34} \\
    \hline
    \end{tabular}
    \begin{flushleft}
    {\small Demographic characteristics and label distribution across four neuroimaging datasets.
    For PPMI, prodromal and SWEDD (Scans Without Evidence of Dopamine Deficit) cases were grouped into HC.
    Abbreviations: HCP, Human Connectome Project; PPMI, Parkinson's Progression Markers Initiative; ABIDE, Autism Brain Imaging Data Exchange; ADNI, Alzheimer's Disease Neuroimaging Initiative; HC, healthy control; MCI, mild cognitive impairment; AD, Alzheimer's disease; PD, Parkinson's disease; W, White; B, Black; A, Asian; I, Indian/Alaska Native; H, Native Hawaiian; O, Other/Multirace; N/A, not available.
    
}
   
    \end{flushleft}
\end{table}

\begin{table}[h!]
    \centering
    \captionsetup{name=Extended Data Table}
    \caption{\textbf{Clinical Dataset Study Population}}
    \label{tab:clinical_population}
    \setlength{\tabcolsep}{10pt} 
    \setlength{\extrarowheight}{8pt}
    \small
    \begin{tabular}{m{1.8cm} m{2.4cm} m{0.6cm} m{1.8cm} m{1.8cm} m{3.7cm}}
    \hline
    \rowcolor[gray]{0.9}
    \textbf{Dataset} & \textbf{Source} & \textbf{n} & \textbf{Age (y)} & 
    \textbf{\makecell{Male n (\%)}} & \textbf{\makecell{Target n (GPi/STN)}} \\
    \hline
    Retrospective TI & \makecell{Ruijin Hospital \\ (Shanghai)} & 51 & 69.84 ± 6.54 & 29 (56.9) & 24/27 \\
    \hline
    Retrospective DBS & \makecell{Ruijin Hospital \\ (Shanghai)} & 55 & 64.05 ± 8.27 & 37 (67.3) & 28/27 \\
    \hline
    Prospective TI & \makecell{Ruijin Hospital \\ (Shanghai)} & 11 & 60.27 ± 8.60 & 9 (81.8) & 4/7 \\
    \hline
    External TI & \makecell{Zhongnan Hospital \\ (Wuhan)} & 14 & N/A & 8 (57.1) & 2/12 \\
    \hline
    \end{tabular}
    \begin{flushleft}
    {\small Demographic and clinical characteristics of the real-world Parkinson's disease cohorts used for fine-tuning and validation. NA, not available.
}
    \end{flushleft}
\end{table}

\end{document}